\documentclass[onecolumn, showpacs,linenumbers]{revtex4}
\usepackage{graphicx}
\usepackage{epstopdf}
\usepackage{caption}
\usepackage{graphicx}
\usepackage{bm}
\usepackage[latin1]{inputenc}
\usepackage{amsmath}
\usepackage{amsfonts}
\usepackage{subfigure}
\usepackage{amssymb}
\usepackage{makeidx}
\usepackage{color}
\usepackage{adjustbox}
\usepackage{lipsum}
\usepackage{mathrsfs}
\usepackage{amsfonts}
\RequirePackage[colorlinks,citecolor=blue,urlcolor=blue,linkcolor=magenta]{hyperref}

\begin{document}

\title{Extended Phase Space Analysis of Interacting Dark Energy Models in Loop Quantum Cosmology}

\smallskip

\author{\bf~ Hmar~Zonunmawia$^1$\footnote{zonunmawiah@gmail.com}, Wompherdeiki Khyllep$^2$\footnote{sjwomkhyllep@gmail.com}, Nandan~Roy$^{3,4}$\footnote{nandan@fisica.ugto.mx}, Jibitesh Dutta$^{5,6}$\footnote{jdutta29@gmail.com,~jibitesh@nehu.ac.in} and Nicola Tamanini$^7$\footnote{nicola.tamanini@cea.fr}}
\smallskip

\affiliation{$^1$Department of Mathematics,~ North Eastern Hill
University,~NEHU Campus, Shillong - 793022 ( INDIA )}

\affiliation{$^{2}$ Department of Mathematics, St. Anthony's College, Shillong, Meghalaya 793001, India}

\affiliation{$^3$ Departamento de F\'isica, DCI Campus
 Le\'on  Universidad de Guanajuato, 37150 Le\'on, Guanajuato, M\'exico}

\affiliation{$^4$ Harish-Chandra Research Institute, Chhatnag Road, Jhunsi, Allahabad-211019, India}

\affiliation{$^{5}$ Mathematics Division, Department of Basic
Sciences and Social Sciences,~ North Eastern Hill University,~NEHU
Campus, Shillong - 793022 (INDIA)}

\affiliation{$^6$ Inter University Centre for Astronomy and Astrophysics, Pune 411 007, India}

\affiliation{$^7$Institut de Physique Th\'eorique, CEA-Saclay,
CNRS UMR 3681, Universit\'e Paris-Saclay, F-91191 Gif-sur-Yvette, France }

\date{\today}

\begin{abstract}

The present work deals with the dynamical system investigation of interacting dark energy models (quintessence and phantom) in the
framework of Loop Quantum Cosmology by taking into account a broad class of self-interacting scalar field potentials.
The main reason for studying potentials beyond the exponential type is to obtain additional critical points which can yield more interesting cosmological solutions.
The stability of critical points and the asymptotic behavior of the phase space are analyzed using dynamical system tools and numerical techniques.
We study two class of interacting dark energy models and consider two specific potentials as examples: the hyperbolic potential and the inverse power-law potential.
We found a rich and interesting phenomenology including the avoidance of big rip singularities due to loop quantum effects, smooth and non-linear transitions from matter domination to dark energy domination and finite periods of phantom domination with dynamical crossing of the phantom barrier.

\end{abstract}

 \maketitle


\section{Introduction}

The accelerated expansion of our universe is by now confirmed by several observations, e.g.~Cosmic Microwave Background (CMB) anisotropies \cite{Bennett:2003bz,Spergel:2003cb,Masi:2002hp,Ade:2013zuv,Ade:2015xua}, large scale galaxy surveys \cite{Scranton:2003in} and type Ia supernovae \cite{Riess:1998cb,Tonry:2003zg,Perlmutter:1998np}, but the nature of the entity that causes it, named dark energy (DE), is still obscure.
The cosmological constant represents the most simple and popular choice as a candidate for DE, but it is troubled by different theoretical issues: in particular the cosmological constant problem and the coincidence problem \cite{Weinberg:1988cp,Martin:2012bt}.
In order to find alternative explanations for the observed acceleration of the universe, there are in general two different approaches one can follow: modified gravity models and dynamical DE models.
Within the modified gravity framework, the underlying gravitational theory determining the cosmological evolution is different from general relativity, while dynamical DE models do not modify the gravitational interaction, but rather introduce a new type of exotic matter component in the universe to describe the accelerated expansion.
In both approaches the cosmic dynamics can often be effectively described by the action of a single scalar field.
Several dynamical scalar field models have been proposed and studied: examples are quintessence \cite{Caldwell:1997ii,Sahni:2002kh,Carroll:1998zi}, phantom DE \cite{Caldwell:1999ew,Singh:2003vx,Elizalde:2004mq} and $k$-essence \cite{ArmendarizPicon:2000ah}.
Scalar field models with self-interacting potentials can provide a useful cosmological evolution which mimics the effect of a cosmological constant at the present epoch.
In a late-time cosmological context, a canonical scalar field is commonly known as quintessence and
can be motivated by the low energy limit of some well known high energy theories, e.g.~string theory.
The quintessence equation of state $w$ (EoS) can take values in the range $-1\leq w \leq 1$, implying accelerated expansion for $w<-1/3$.
A canonical scalar field however cannot produce the so-called phantom regime ($w<-1$) which is slightly favoured by astronomical observations \cite{Novosyadlyj:2013nya}.
For this reason, another well-known scalar field model of DE has been proposed: the phantom field with negative kinetic energy.
Phantom fields are plagued by instabilities at the quantum level \cite{Carroll:2003st}, but if considered from an effective phenomenological perspective they can be used as interesting cosmological solutions, which may better fit the observational data.

In standard Einstein cosmology (EC), phantom DE models usually lead to a cosmic end described by a future big rip singularity \cite{Caldwell:2003vq}.
This behaviour is however expected to be corrected by loop quantum effects.
Loop Quantum Gravity (LQG) \cite{cr} is one of the well-known approaches to a quantum theory of gravity \cite{Han:2005km}.
Its main aim consists in quantizing gravity with a non-perturbative and background independent method \cite{Thiemann:2002nj,Ashtekar:2003hd}.
The application of LQG in the context of cosmology is called Loop Quantum Cosmology (LQC) \cite{Ashtekar:2006rx}.
LQC modifies the standard Friedmann equation by adding a term depending on a fix energy density imposed by quantum corrections
\cite{Bojowald:2001ep,Date:2004zd,Ashtekar:2006uz,Banerjee:2005ga}, which essentially encodes the discrete quantum geometric nature of spacetime \cite{Ashtekar:2006wn}.
This modifying contribution in the standard Friedmann equation can be used to avoid any past and future singularity \cite{Ashtekar:2006rx,Singh:2006sg,Copeland:2005xs}.
In fact loop corrections become important when the total energy density of the universe approaches the critical high energy value predicted by the theory (cf.~Eq.~\eqref{eq:rho_c}), in which case a cosmic bounce might occur and the big rip and other singularities will never be reached \cite{Ashtekar:2006wn,Sami:2006wj,Naskar:2007dn,Samart:2007xz}.
In both EC and LQC when DE is modelled as a scalar field, this is usually assumed to interact only with itself.
However there is no fundamental argument to ignore a possible coupling between DE and dark matter (DM), and DE models where a scalar field interacts non-gravitationally with the matter sector have been proposed as well; see e.g.~\cite{Amendola:1999er,Billyard:2000bh,Chimento:2003iea,Cai:2004dk,Huey:2004qv,Boehmer:2008av,Boehmer:2009tk}.
Interacting DE models are known to produce late time accelerated scaling attractors which can be used to alleviate the cosmic coincidence problem \cite{Chimento:2003iea}.
However since there is no experimental evidence for a dark sector interaction and the fundamental nature of both DE and DM is still obscure, any coupling between the two dark components is phenomenologically constructed at the level of the equations of motion, though some recent attempts to built an effective interaction at the Lagrangian level have been advanced \cite{Pourtsidou:2013nha,Boehmer:2015kta,Boehmer:2015sha,Koivisto:2015qua}.

In this work, using dynamical system tools, we investigate the dynamics of interacting scalar field (quintessence and phantom) in the framework of LQC for a broad class of self-interacting potentials.
Similar studies, restricted to the exponential potential case, have already been performed \cite{Samart:2007xz,Gumjudpai:2007fc,Fu:2008gh}, and the analysis of different scalar field potential will allow us to better understand the complete cosmological potential of these DE models, at least at the background level.
This type of generalization has been widely studied in different cosmological frameworks: e.g.~standard quintessence models \cite{Roy:2014yta,Paliathanasis:2015gga}, braneworld theories \cite{Escobar:2013js,Dutta:2015jaq,Dutta:2016dnt}, $k$-essence \cite{Dutta:2016bbs}, chameleon theories \cite{Roy:2014hsa}, scalar-fluid theories \cite{Dutta:2017kch} and (non-interacting) LQC \cite{Xiao:2011nh}.
In these investigations the dimension of the resulting dynamical system increases by one if compared to that of the exponential potential case, making the analysis slightly more complicated.
The analysis of other scalar field potentials, beyond the exponential one, helps to better relate these phenomenological models with more fundamental high-energy theories.
Moreover, from a mathematical point of view, these generalizations usually yield additional non-hyperbolic points where linear stability theory fails and the stability properties can only be determined analytically using center manifold theory or Lyapunov functions \cite{Dutta:2017kch,Coley,swig,lper}.
One can alternatively use numerical methods, for example analysing the behaviour of perturbed trajectories near the non-hyperbolic critical point \cite{Roy:2014hsa,Dutta:2016dnt,Dutta:2016bbs}, while for the case of normally hyperbolic points, namely a non-isolated set of critical points with one vanishing eigenvalue, their stability is determined by the signature of the remaining non-vanishing eigenvalues \cite{Roy:2014hsa,Dutta:2016dnt}.
For a better understanding of the cosmological dynamics of models presenting non-hyperbolic critical points, in what follows we consider two concrete potentials as examples: the hyperbolic potential $V=V_0\,\cosh^{-\mu}(\lambda\phi)$ and the inverse power-law potential $ V=M^{4+n}/\phi^n$.

In our analysis we consider two interacting models based on two different coupling functions for both the quintessence and phantom fields.
The first one arises in string theory and it has already been studied in the case of standard EC (for both quintessence and phantom case) \cite{Boehmer:2008av}.
The same interaction has also been studied in the LQC framework for a phantom field with exponential potential \cite{Fu:2008gh}, showing that in such case the big rip singularity can be avoided.
The second kind of interaction was studied recently for quintessence in the EC context \cite{Shahalam:2015sja}, showing that the coincidence problem can be alleviated.
It was also investigated in braneworld theories with both quintessence and phantom fields as DE \cite{Dutta:2016dnt}.
In both interactions, depending on the choice of the scalar field potential, we find that there is a late time attractor with
contribution from loop quantum gravity.

The paper is organized as follows.
Section~\ref{section2} deals with the basic equations of interacting dark energy in LQC and shows how they can be recast into an autonomous system of equations.
In Sections~\ref{section3} and \ref{section4} we consider two interacting dark energy models and we investigate the corresponding cosmological evolution using dynamical system tools.
In each of these section, we present two subsections: one for quintessence and the another for phantom DE, wherein we consider two scalar field potentials as examples.
The last two sections~\ref{sec:cosmological_implications} and \ref{section5} are devoted to discussing the cosmological implications and drawing conclusions, respectively.


\section{Dynamics of interacting scalar field dark energy in Loop Quantum Cosmology}
\label{section2}

In a flat universe the effective modified Friedmann equation in the framework of LQC is given by \cite{Bojowald:2001ep,Singh:2006sg}
\begin{equation}\label{1}
3H^2= \frac{8 \pi G}{c^4} \rho\Big(1-\frac{\rho}{\rho_c}\Big),
\end{equation}
where $H$ is the Hubble parameter, $\rho=\rho_m+\rho_\phi$ is the total energy density and $\rho_\phi$, $\rho_m$ are the energy densities of dark energy and dark matter respectively.
The constant
\begin{equation}
  \rho_c=\frac{\sqrt{3}}{16 \pi^2\gamma^3\,G^2\hbar} \,,
  \label{eq:rho_c}
\end{equation}
is the critical loop quantum density, where $\gamma$ is the dimensionless Barbero-Immirzi parameter \cite{Bojowald:2001ep,Date:2004zd,Ashtekar:2006uz,Banerjee:2005ga}.
To simplify the notation in what follows we shall use units where $8 \pi G \equiv c \equiv 1$.
Although the single energy components $\rho_m$ and $ \rho_{\phi}$ may not be conserved separately, the total energy density is conserved
\begin{equation} \label{con}
\dot{\rho} + 3 H ( \rho + p) = 0 \,,
\end{equation}
where $p$ is the total pressure and an over-dot denotes differentiation with respect to the time $t$.
One can also write the modified Raychaudhuri equation of the system as
\begin{equation} \label{raychau}
\dot{H} = - \frac{1}{2} (\rho + p) (1 - 2 \frac{\rho}{\rho_c}) \,.
\end{equation}
We assume that DE is described by either quintessence or a phantom scalar field.
The general Lagrangian for both these scalar fields can be generally written as
\begin{equation} \label{lagrangian}
\mathcal{L} = \frac{1}{2} ~ \epsilon ~ \partial^{\mu} \phi \partial_{\mu} \phi - V(\phi) \,,
\end{equation}
where $\epsilon = 1$ corresponds to quintessence and $\epsilon = -1$ corresponds to the phantom field.
Here $V(\phi)$ is the self-interacting potential for the scalar field $\phi$.
The energy density and pressure of the scalar field are respectively given by
\begin{equation}\label{2}
\rho_{\phi}= \epsilon~\frac{1}{2}\dot{\phi}^2 +V(\phi)
\hspace{1.5cm} \text{and} \hspace{1.5cm} p_{\phi}=\epsilon~\frac{1}{2}\dot{\phi}^2  -V(\phi).
\end{equation}
In our investigation we assume that the scalar field interacts with dark matter and the energy conservation equations take the form
\begin{align}
  \dot{\rho_\phi}+3H(1+w_\phi)\rho_\phi &= -Q \,,\label{3}\\
  \dot{\rho_m}+3H\rho_m &= +Q \,, \label{4}
\end{align}
where $Q$ is the interaction term and $w_\phi = p_\phi/\rho_\phi$ is the EoS of DE.
If $Q$ is positive, then the energy transfer takes place from quintessence/phantom DE to DM, whereas for a negative $Q$ energy flows from DM to DE.
From Eqs.~(\ref{2}) and (\ref{3}) the evolution equation of the scalar field can be expressed as
  \begin{equation}\label{5}
   {\ddot{\phi}}=-3H\dot\phi-\frac{1}{\epsilon}\frac{dV}{d\phi}-\frac{Q}{\epsilon\,\dot{\phi}} \,,
   \end{equation}
while from Eq.~(\ref{raychau}) the effective modified Raychaudhuri equation can be rewritten as
\begin{equation}\label{6}
\dot{H}=-\frac{1}{2}\Big(\rho_m+\epsilon\dot{\phi}^2\Big)\Big(1-\frac{2\rho}{\rho_c}\Big) \,.
\end{equation}

In order to write our system of equations as an autonomous system of ordinary differential equations, we introduce the following set of dimensionless phase space variables
  \begin{equation}\label{7}
   x\equiv \frac{\dot\phi}{\sqrt{6}~H} , \hspace{0.2cm}  y\equiv\frac{\sqrt{V(\phi)}}{\sqrt{3}~H} \hspace{0.2cm},
    \hspace{0.2cm} z\equiv\frac{\rho}{\rho_c} \hspace{0.2cm} \textrm{and} \hspace{0.2cm} s\equiv -\frac{1}{V}\frac{dV}{d\phi}.
   \end{equation}
Using the new dimensionless variables (\ref{7}), the effective modified Friedmann equation (\ref{1}), can be rewritten in a dimensionless form as
   \begin{equation}\label{8}
   1 =\Big(\frac{\rho_m}{3H^2}+\epsilon x^2+y^2\Big)(1-z) \,.
   \end{equation}
This provides a constraint that can be used to eliminate $\rho_m$ in favour of the dimensionless variables (\ref{7}) in all equations that follow.
Using equations Eqs.~\eqref{5}--\eqref{7}, we obtain the following autonomous system of differential equations,
\begin{align}
 x'&=-3x+\frac{1}{\epsilon}\sqrt{\frac{3}{2}}~sy^2+x\Big[\frac{3}{2}\Big(\frac{1}{1-z}-\epsilon x^2-y^2\Big)+\epsilon3x^2\Big](1-2z)-\frac{Q}{\epsilon~\sqrt{6}~H^2\dot\phi},\label{x_prime}\\
y'&=-\sqrt{\frac{3}{2}}~sxy +y\Big[\frac{3}{2}\Big(\frac{1}{1-z}-\epsilon x^2-y^2\Big)+\epsilon3x^2\Big](1-2z),\label{y_prime}\\
z'&=-3z-3z(1-z)(\epsilon x^2-y^2),\label{z_prime}\\
s'&=-\sqrt{6}~xf(s),\label{s_prime}
\end{align}
where a prime denotes differentiation with respect to $N=\ln a$ ($a$ being the usual scale factor) and we defined
\begin{equation}
  f(s)=s^2(\Gamma(s)-1) \qquad\mbox{and}\qquad \Gamma= V\frac{d^2V}{d\phi^2}\Big({\frac{dV}{d\phi}}\Big)^{-2} \,.
  \label{eq:gamma}
\end{equation}
We consider a class of potentials where $\Gamma$ is a function of $s$, with $\Gamma=1$ being the case of an exponential potential.
In general $\Gamma$ may not be function of $s$, in which case one has to take into account the higher derivatives of the scalar field potential \cite{Xiao:2011nh} or choose new dimensionless variables \cite{Lazkoz:2007mx,Nunes:2000yc}, resulting in the increase of dimension of the dynamical system for both cases.
In terms of the dimensionless variables (\ref{7}), the scalar field relative energy density, the relative energy density of dark matter, the relative energy contribution due to loop quantum corrections, the total cosmic energy EoS and the deceleration parameter are given by
  \begin{align}
  \Omega_\phi&\equiv \frac{\rho_{\phi}}{3H^2}=\epsilon x^2+y^2,\label{11}\\
   \Omega_m&\equiv \frac{\rho_{m}}{3H^2}=\frac{1}{1-z}-\epsilon x^2-y^2,\label{12}\\
   \Omega_c & \equiv 1 - \Omega_m - \Omega_\phi = \frac{z}{z-1},\label{13}\\
  w_{\rm tot}&\equiv \frac{p_{\phi}}{\rho_{\phi}+\rho_m}=(1-z)(\epsilon x^2-y^2),\label{14}\\
 q&\equiv -1-\frac{\dot{H}}{H^2}=-1+\Big[\frac{3}{2}\Big(\frac{1}{1-z}-\epsilon x^2-y^2\Big)+\epsilon3x^2\Big](1-2z). \label{15}
 \end{align}
The dynamical systems analysis yields the asymptotic behavior of the system, i.e.,~the beginning and the ultimate fate of the universe. For the numerical calculations that follows, we choose initial conditions in such a way that the final state of the universe is in agreement with the present observational data: $\Omega_m = 0.31$, $\Omega_\phi = 0.69$ and $q=-0.62$ \cite{Ade:2015xua}.

In order to close the autonomous system \eqref{x_prime}--\eqref{s_prime}, we need to consider a particular form of the interaction $Q$ for which the last term in Eq.~\eqref{x_prime} can be expressed as a function of the variables \eqref{7}.
In the next two sections, we choose the following two particular interaction terms
\begin{center}
$ (I)~Q=\alpha\rho_m\dot\phi$~~~~~~~~ $ (II)~Q=\beta\dot\rho_\phi$ \,,
\label{eq:couplings}
\end{center}
with $\alpha$ and $\beta$ dimensionless constants.
The first of these couplings arises naturally from scalar-tensor theories in the Einstein frame \cite{Wetterich:1994bg,Amendola:1999er,Holden:1999hm} and is well motivated from string theory. The second one is instead purely phenomenological \cite{Shahalam:2015sja} and it will be used to expose some interesting properties of a more complex interacting model.
In what follows, we tried to keep the analysis as general as possible but in order to study the stability of some interesting critical points in more detail, we consider two particular forms for the scalar field potential as examples: $V=V_0\,\cosh^{-\mu}(\lambda\phi)$ and $ V=M^{4+n}/\phi^n$.
We note also that in non-interacting DE models, the energy density of DM is always taken to be non-negative, while in models of interacting DE the condition $\Omega_m<0$ can be allowed (see e.g.~\cite{Quartin:2008px}).
In particular this means that the resulting phase spaces are generally not compact.
This implies that critical points at infinity should be analyzed by compactifying the phase space using the Poincare compactification technique; see e.g.~\cite{Tamanini:2014nvd}.
Nevertheless in what follows we only determine the dynamics near the finite critical points, which is enough from a phenomenological point of view since our aim is to find physically viable solutions, namely trajectories connecting DM to DE domination.
Once we find such type of solutions, then given the right initial conditions the observed evolution of our universe can be described by the DE model under consideration (at least at the background level).


\section{Interacting model I: $\bf Q=\alpha\rho_m\dot\phi$}
\label{section3}

This type of interaction term arises naturally in scalar-tensor theory \cite{Wetterich:1994bg,Amendola:1999qq}, where the energy terms are separately conserved in the Jordan frame but become coupled in the Einstein frame.
The dynamical system investigation of quintessence with this interaction and an exponential potential has been studied in both standard EC \cite{Boehmer:2008av} and LQC \cite{Samart:2007xz,Gumjudpai:2007fc,Fu:2008gh}.
In this section, we provide the dynamical analysis of this DE model with a general scalar field potential in the LQC framework.
Here we only show the properties of the phase space and characterize the full cosmological dynamics of this model.
Discussions about the cosmological implications are postponed to section~\ref{sec:cosmological_implications}.

As noted above the dark sector interaction modifies only the $x^\prime$ equation in the system (\ref{x_prime})-(\ref{s_prime}), while the other equations remain unaffected.
For this particular interaction the system of equations (\ref{x_prime})-(\ref{s_prime}) becomes
\begin{align}
x'&=-3x+\frac{1}{\epsilon~}\sqrt{\frac{3}{2}}~sy^2 -\frac{\sqrt{6}}{\epsilon~2}\alpha\Big(\frac{1}{1-z}-\epsilon x^2-y^2\Big)+x\Big[\frac{3}{2}\Big(\frac{1}{1-z}-\epsilon x^2-y^2\Big)+\epsilon3x^2\Big](1-2z),\label{x_I}\\
y'&=-\sqrt{\frac{3}{2}}~sxy +y\Big[\frac{3}{2}\Big(\frac{1}{1-z}-\epsilon x^2-y^2\Big)+\epsilon3x^2\Big](1-2z),\label{y_I}\\
z'&=-3z-3z(1-z)(\epsilon x^2-y^2),\label{z_I}\\
s'&=-\sqrt{6}~xf(s)\,.\label{s_I}
\end{align}
It is easy to check that the above dynamical system (\ref{x_I})-(\ref{s_I}) is invariant under the transformation $y\rightarrow -y$, meaning that we need to analyze only the phase space for positive values of $y$.
The critical points of the system (\ref{x_I})-(\ref{s_I}) and the relevant parameters are listed in Table \ref{Table1} and the eigenvalues of the corresponding Jacobian matrix are listed in Table~\ref{Table2}.
In these tables, and throughout the rest of the paper, $s_*$ is a solution of the equation $f(s)=0$ and $df(s_*)$ denotes the value of the derivative of $f$ at $s=s_*$.
From Tables~\ref{Table1} and \ref{Table2}, we can note that all critical points depend on the specific form of potentials: $C_1$, $C_2$, $C_3$, $C_4$, $C_5$ all depend on $s_*$ and $df(s_*)$, whereas $C_6$ depends on the value of $f(0)$ for its stability.
We now discuss the stability of the critical points listed in Table~\ref{Table1} for quintessence and phantom dark energy separately.

\begin{table}
\centering
\begin{adjustbox}
{width=1\textwidth}
\small

\begin{tabular}{|c|c|c|c|c|c|c|c|c|c|c|}
\hline
Point&$x$~&$y$~& $z$~&$s$~&${\Omega}_{\phi}~$&${\Omega}_{m}~$&$q$~&$w_{\rm tot}$~&Existence \\\hline
\hline

$C_1$&$\frac{1}{\sqrt{\epsilon}}$~~~&$0$~&$0$~&$s_*$~&$1$~&$0$~&$2$~&$1$~&$\epsilon=1$ \\[2ex]
$C_2$&$-\sqrt{\frac{2}{3}}~\frac{\alpha}{\epsilon}$~&$0$~&$0$~&$s_*$~&$\frac{2\alpha^2}{3\epsilon}$~&$1-\frac{2\alpha^2}{3\epsilon}$~&$\frac{1}{2}\frac{2\alpha^2+\epsilon}{\epsilon}$~&$\frac{2\alpha^2}{3\epsilon}$&Always\\
$C_3$&$\sqrt{\frac{3}{2}}~\frac{1}{\alpha}$~&$0$~&$\frac{2\alpha^2+3\epsilon}{3\epsilon}$~&$s_{*}$~&$\frac{3\epsilon}{2\alpha^2}$~&$-\frac{3\epsilon}{\alpha^2}$~&$-1$&$-1$ ~&$\alpha \neq 0$\\[2ex]
$C_4$&$\frac{s_*}{\sqrt{6}\epsilon}$~&$\sqrt{1-\frac{s^2_*}{6\epsilon}}$~&$0$~&$s_*$~&$1$~&$0$~&$-1+\frac{s_*^2}{4}\epsilon(\epsilon^2+1)$~&$-1+\frac{s_*^2}{6}\epsilon(\epsilon^2+1)$~&Always for $\epsilon=-1$\\&&&&&&&&&$s_*^2<6$ for $\epsilon=1$\\[2ex]
$C_5$&$\sqrt{\frac{3}{2}}~\frac{1}{(\alpha+s_*)}$~&$\frac{\sqrt{2\alpha^2+3\epsilon+2\alpha s_*}}{\sqrt{2}(\alpha+s_*)}$~&$0$~&$s_*$~&$\frac{\alpha^2+\alpha s_*+3\epsilon}{(s_*+\alpha)^2}$~&$\frac{s^2_*+\alpha s_*-3\epsilon}{(s_*+\alpha)^2}$~&$-\frac{1}{2}\frac{2\alpha-s_*}{\alpha+s_*}$&$-\frac{\alpha}{\alpha+s_*}$~&$2\alpha(\alpha+s_*)+3\epsilon>0$ \\
$C_6$&$0$~&$\frac{1}{\sqrt{1-z}}$~&$z$~&$0$~&$\frac{1}{1-z}$~&$0$~&$-1$~&$-1$~&$z<1$
\\\hline
\end{tabular}
\vspace{0.5cm}
\end{adjustbox}
\caption{Critical points of the system (\ref{x_I})-(\ref{s_I}) and values of the relevant cosmological quantities for a generic scalar field potential.} \label{Table1}
\end{table}

\begin{table}
\centering

Here: $\mu_{1\mp}=-\frac{3}{4}\frac{2\alpha+s_*}{\alpha+s_*}\Big(1+\sqrt{1+\frac{8\left[3\pm s_*(s_*+\alpha)\right]
\left[2\alpha^2+2\alpha\,s_*\mp 3\right]}{3\left[2\alpha+s_*\right]^2}}\Big)$,
$\mu_{2\mp}=-\frac{3}{4}\frac{2\alpha+s_*}{\alpha+s_*}\Big(1-\sqrt{1+\frac{8\left
[3\pm s_*(s_*+\alpha)\right]\left[2\alpha^2+2\alpha\,s_*\mp 3\right]}{3\left[2\alpha+s_*\right]^2}}\Big)$

\begin{adjustbox}{width=1\textwidth}

\small

\begin{tabular}{|c|c|c|c|c|c|c|}

\hline
Point & $\lambda_1$~ & $\lambda_2$~ & $\lambda_3$~ & $\lambda_4$&Stability \\\hline
\hline
&&&&& \\
$C_1$&$-6$~&$3-\frac{s_*}{2}\sqrt{\frac{6}{\epsilon}}$~&$3+\sqrt {6}\alpha\frac{1}{\sqrt{\epsilon}}$~&$-\sqrt{\frac{6}{\epsilon}}\,df(s_*)$&$\alpha<-\frac{\sqrt{6}}{2}$, $s_*>\sqrt{6}$, $df(s_*)>0\,$ (See Fig. \ref{fig:c5_I_quint_region} for $\epsilon=1$)\\
&&&&& Saddle: otherwise\\[3ex]
&&&&&See Fig. \ref{fig:c5_I_quint_region} for $\epsilon=1$\\
$C_2$&$-\Big({\frac {2\,{\alpha}^{2}+3\,\epsilon }{\epsilon }}\Big)$&$\frac{1}{2}\Big({\frac {2\,{\alpha}^{2}-3\,\epsilon }{\epsilon }}\Big)$~&$\frac{1}{2}\Big(\frac{2\alpha^2+2\alpha s_{*}+3\epsilon}{\epsilon}\Big)$~&$\frac{2\alpha df(s_{*})}{\epsilon}$& See Fig. \ref{fig:c5_I_phan_region} for $\epsilon=-1$ \\
&&&&& Saddle: otherwise \\[3ex]
&&&&&$\alpha\,s_*>0$, $\alpha\,df(s_*)>0$ (See Fig. \ref{fig:c5_I_quint_region} for $\epsilon=1$)\\
$C_3$&$-\frac{3s_{*}}{2\alpha}$~&$-\frac{3}{2}\Big({\frac {\alpha-\sqrt {-3\,{\alpha}^{2}-6\,\epsilon }}{\alpha}}\Big)$~&$-\frac{3}{2}\Big({\frac {\alpha+\sqrt {-3\,{\alpha}^{2}-6\,\epsilon }}{\alpha}}\Big)$&$-\frac{3\,df(s_{*})}{\alpha}$& ~~~ See Fig. \ref{fig:c5_I_phan_region} for $\epsilon=-1$\\
&&&&& Saddle: otherwise \\[3ex]
$C_4$&$-\frac{s^2_*}{\epsilon}$~&$\frac{1}{2}\Big(\frac{s^2_*-6\epsilon}{\epsilon}\Big)$~&$\frac{\alpha s_*+s^2_*-3\epsilon}{\epsilon}$~&$\frac{s_*df(s_*)}{\epsilon}$&$s_*(\alpha+s_*)<3$, $s_*df(s_*)<0$  (See Fig. \ref{fig:c5_I_quint_region} for $\epsilon=1$)~\\&&&&&Saddle ($\epsilon=-1$)\\[2ex]
$C_5$&$-\frac{3s_*}{\alpha+s_*}$~&$\mu_{1\mp}$~&$\mu_{2\mp}$~&$-\frac{3\,df(s_*)}{\alpha+s_*}$&See Fig. \ref{fig:c5_I_quint_region} for $\epsilon=1$\\
 &&&&&Saddle ($\epsilon=-1$)\\[2.5ex]
$C_6$&$0$~&$-3$~&$\frac{3}{2}\left(-1+\sqrt{\left(1+\frac{4\,f(0)}{3\epsilon(z-1)}\right)}\right)$~&$\frac{3}{2}\left(-1-\sqrt{\left(1+\frac{4\,f(0)}{3\epsilon(z-1)}\right)}\right)$~~&$f(0)>0~ (\epsilon=1)$\\&&&&&$f(0)<0~ (\epsilon=-1)$\\
\hline

\end{tabular}

\vspace{0.5cm}

\end{adjustbox}
\caption{Eigenvalues of the Jacobian matrix of the system (\ref{x_I})-(\ref{s_I}). Here $\mu_{1+}$, $\mu_{2+}$ correspond to $\epsilon=1$ while $\mu_{1-}$, $\mu_{2-}$ correspond to $\epsilon=-1$.} \label{Table2}
\end{table}
\vspace{.5cm}

\subsection{\bf Quintessence dark energy $(\epsilon=1)$}

\begin{figure}
\centering
\subfigure[]{%
\includegraphics[width=6cm,height=4cm]{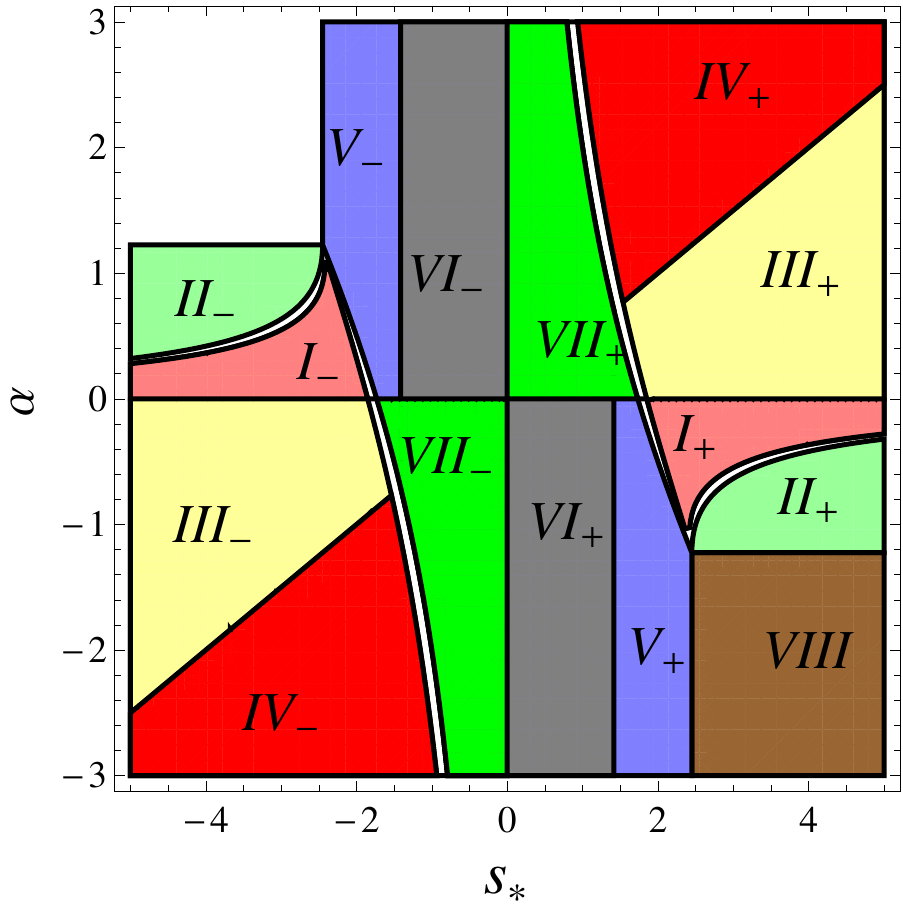}\label{fig:c5_I_quint_region}}
\qquad
\subfigure[]{%
\includegraphics[width=6cm,height=4cm]{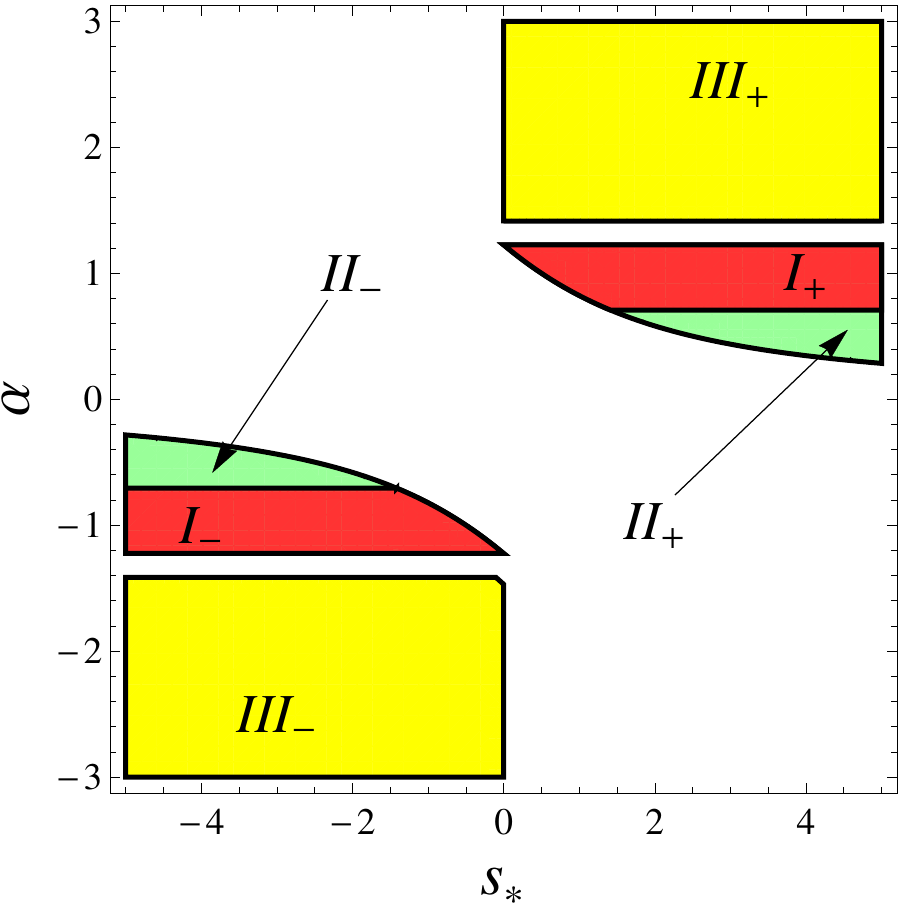}\label{fig:c5_I_phan_region}}
\caption{
Stability regions of points $C_1$, $C_2$, $C_3$, $C_4$, $C_5$ in the $(s_*,\alpha)$ parameter space.
In panel (a) we consider $\epsilon=1$ and in panel (b) $\epsilon=-1$.
\textbf{In panel (a)}: Region $VIII$ represents the region of stability of point $C_1$ for potentials with $df(s_*)>0$, regions $II_+$ and $II_-$ represent the regions of stability of point $C_2$ for potentials with $df(s_*)>0$ and $df(s_*)<0$, respectively. Regions $III_+$, $IV_+$, $VII_+$ represent the regions of stability of point $C_3$ for potentials with $df(s_*)>0$, whereas regions $III_-$, $IV_-$, $VII_-$ represent the regions of stability of point $C_3$ for potentials with $df(s_*)<0$. Regions $V_+$, $VI_+$, $VII_+$ represent the regions of stability of point $C_4$ for potentials with $df(s_*)<0$, whereas regions $V_-$, $VI_-$, $VII_-$ represent the regions of stability of point $C_4$ for potentials with $df(s_*)>0$. Regions $I_+$, $III_+$, $IV_+$ represent the regions of stability of point $C_5$ for potentials with $df(s_*)>0$, whereas regions $I_-$, $III_-$, $IV_-$ represent the regions of stability of point $C_5$ for potentials with $df(s_*)<0$. Regions $VI_+$, $VI_-$,$VII_+$, $VII_-$ denote the regions of acceleration of point $C_4$, whereas regions $IV_+$, $IV_-$ denote the regions of acceleration of point $C_5$.
\textbf{In panel (b)}: Regions $I_+$, $II_+$ represent the regions of stability of point $C_2$ for potentials with $df(s_*)>0$, whereas regions $I_-$, $II_-$ represent the
 regions of stability of point $C_2$ for potentials with $df(s_*)<0$. Region $III_+$ represents the region of stability of point $C_3$ for potentials with $df(s_*)>0$, whereas
 region $III_-$ represents the region of stability of point $C_3$ for potentials with $df(s_*)<0$. Regions $I_+$, $I_-$ denote the regions of acceleration of point $C_2$.}
\label{fig:stability_region_I}
\end{figure}

We briefly discuss the properties of the critical points listed in Table~\ref{Table1} with $\epsilon=1$.
The stability regions in the $(s_*,\alpha)$ parameter space for each critical point are shown in Fig.~\ref{fig:c5_I_quint_region}.

\begin{itemize}

\item {\bf Point $C_1$}: This point corresponds to a decelerated, stiff matter dominated universe ($w_{\rm tot}=1$). It is stable whenever $\alpha<-\frac{\sqrt{6}}{2}$, $s_*>\sqrt{6}$, $df(s_*)>0$, otherwise it is saddle.

\item {\bf Point $C_2$}: This point corresponds to a decelerated, scaling solution. It is stable whenever $\alpha^2<3$, $\alpha\,s_*<-3$, $\alpha\,df(s_*)<0$, otherwise it is saddle.

\item {\bf Point $C_3$}: This point corresponds to an accelerated solution ($q=-1$) with loop quantum contribution and negative DM energy density ($\Omega_m<0$). It corresponds to a late time attractor (stable spiral) when $\alpha\,s_*>0$ and $\alpha\,df(s_*)>0$.

\item {\bf Point $C_4$}: Point $C_4$ exists when $s_*^2<6$. It corresponds to an accelerated solution for $s_*^2<2$ and describes a scalar field dominated universe. It is stable whenever $s_*(\alpha+s_*)<3$ and $s_*df(s_*)<0$, otherwise it is saddle.

\item {\bf Point $C_5$}: Point $C_5$ corresponds to a late time accelerated, scaling solution for some values of the parameters $\alpha$ and $s_*$. This is confirmed numerically from Fig.~\ref{fig:c5_I_quint_region} by plotting the region of stability and the region of acceleration for $C_5$ in the $(s_*,\alpha)$ parameter space.

\item {\bf Set of points $C_6$}: A set of non-isolated critical points $C_6$ corresponds to the case where $s=0$ (i.e.~the potential is effectively constant). It is a non-hyperbolic set with one vanishing eigenvalue if $f(0) \neq 0$. This type of
non-isolated points form a normally hyperbolic set \cite{Coley}.
The center manifold of a set of non-isolated critical points is
determined by the direction of the eigenvectors corresponding to
vanishing eigenvalues, while the signature of the non-vanishing
eigenvalues determine its stability. For points $C_6$ the real
components of the non-vanishing eigenvalues are negative only if
$f(0)>0$, implying that this set of points corresponds to a late
time stable attractor only if $f(0)>0$. It is a stable node if
$0<f(0)<\frac{3}{4}(1-z)$, while it is a stable spiral if
$f(0)>\frac{3}{4}(1-z)$. In any other case it is saddle. Further
investigation is required when $f(0)=0$, which will be studied for
the examples below once the scalar field potential is specified.
This additional set of points is interesting from a phenomenological perspective as it shows the effect of loop quantum corrections to explain late time acceleration of the universe.

\end{itemize}
The properties of points $C_1$, $C_2$, $C_4$, $C_5$ are the same as in standard EC \cite{Boehmer:2008av}. Point $C_3$ shows the effect of loop quantum corrections but implies a negative DM energy density.
The additional set of points $C_6$ is interesting as it corresponds to a scalar field dominated solution where the effects of loop quantum gravity corrections can be used to explain the late time accelerated universe.
From the above analysis, we note that critical points $C_1$, $C_2$, $C_4$, $C_5$ cannot be late time attractors simultaneously.
However there is a possibility of multiple late time attractors: points $C_3$ and $C_4$ on one side, and $C_3$ and $C_5$ on the other side.
This kind of situation is interesting from both a phenomenological and mathematical point of view, and it usually leads to a more complex cosmological dynamics.
Depending on the choices of parameters and initial conditions, the interacting DE model considered here can successfully describe the late time evolution of the universe.
Since the existence and stability properties of all critical points heavily depend on $s_*$, $df(s_*)$ and $f(0)$, the full dynamics of the phase space can be derived only once a particular scalar field potential has been chosen.
For this reason in what follows we consider two particular potentials.

\begin{figure}
\centering
\subfigure[]{%
\includegraphics[width=6cm,height=4cm]{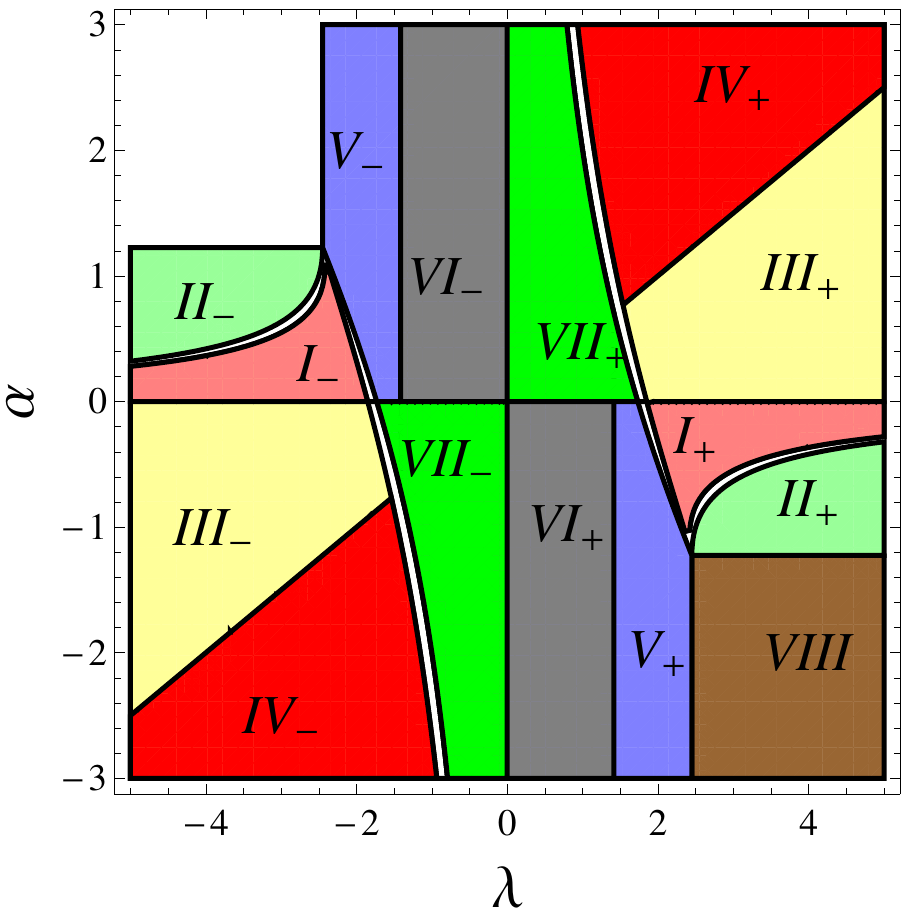}\label{fig:c5_I_quint_region_sinh}}
\qquad
\subfigure[]{%
\includegraphics[width=6cm,height=4cm]{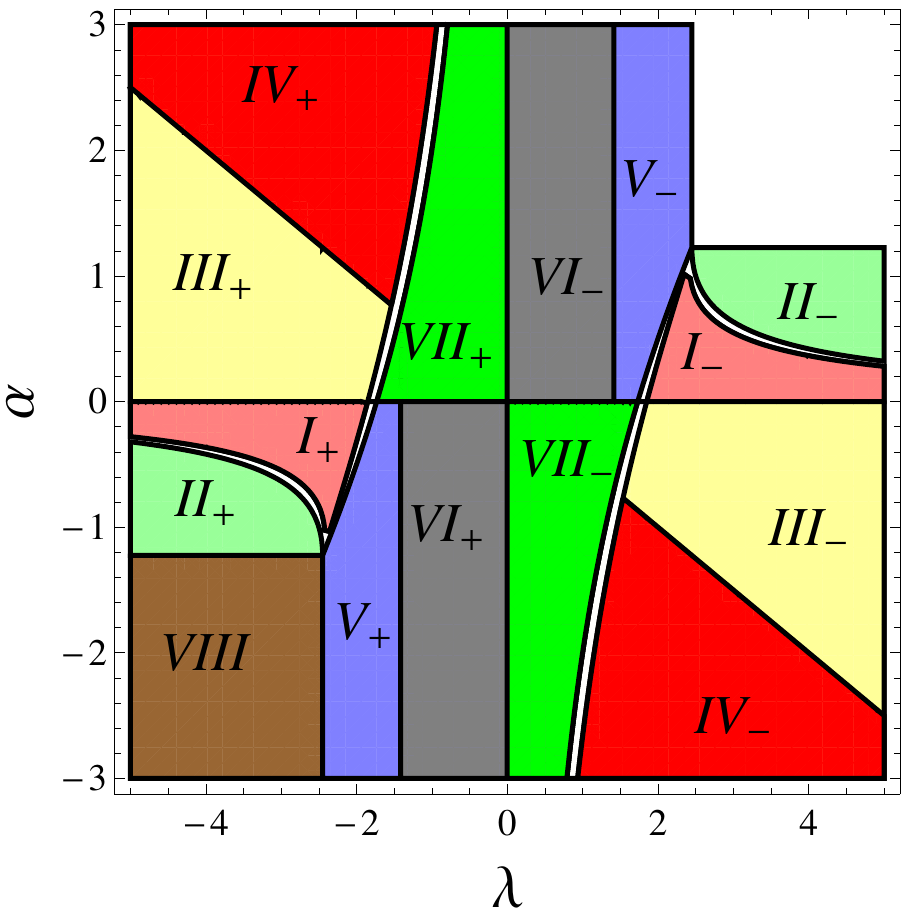}\label{fig:c5_I_quint_region_sinh_neg}}
\caption{(a). Stability regions of points $C_i^+$ ($i=1,2,3,4,5$) in $(\lambda,\alpha)$ parameter space (b). Stability regions of points $C_i^-$ ($i=1,2,3,4,5$) in $(\lambda,\alpha)$ parameter space with potential $V=V_0\,\cosh^{-\mu}(\lambda\phi)$ by taking $\mu=1$. In respective panel, Region $VIII$ represents region of stability of point $C_1^+$ ($C_1^-$), regions $II_+$ and $II_-$ represent regions of stability of point $C_2^+$ ($C_2^-$). Regions $III_+$, $IV_+$, $VII_+$, $III_-$, $IV_-$, $VII_-$ represent regions of stability of point $C_3^+$ ($C_3^-$).  Regions $V_+$, $VI_+$, $VII_+$, $V_-$, $VI_-$, $VII_-$ represent regions of stability of point $C_4^+$ ($C_4^-$). Regions $I_+$, $III_+$, $IV_+$, $I_-$, $III_-$, $IV_-$ represent regions of stability of point $C_5^+$ ($C_5^-$). Regions $VI_+$, $VI_-$,$VII_+$, $VII_-$ represent regions of acceleration of point $C_4^+$ ($C_4^-$), whereas regions $IV_+$, $IV_-$ represent regions of acceleration of point $C_5^+$ ($C_5^-$). Here $\epsilon=1$.}
\label{fig:stability_region_I_sinh}
\end{figure}

\subsubsection*{Example 1: $V=V_0\,\cosh^{-\mu}(\lambda\phi)$} \label{subsec:model_I_quin_cosh}

Here we consider the potential $V=V_0\,\cosh^{-\mu}(\lambda\phi)$, (where $V_0$ and $\lambda$ are two constants of suitable dimensions while $\mu$ is a dimensionless parameter).
This potential was proposed to explain the exit of the universe from a scaling regime to a de-Sitter like accelerated attractor through a mechanism of spontaneous symmetry breaking \cite{Zhou:2007xp}.
The dynamical behaviour of standard EC with this scalar field potential has been studied in \cite{Fang:2008fw}.
For this potential we find $f(s)=\frac{s^2}{\mu}-\mu\lambda^2$, so that
\begin{equation}
 s_*={\pm}\mu\lambda  \hspace{1.5cm}\mbox{and}\hspace{1.5cm} df(s_*)=\frac{2s_*}{\mu} = \pm 2 \lambda \,.
 \end{equation}
Each of the critical points $C_1$, $C_2$, $C_3$, $C_4$, $C_5$
appears twice in the phase space according to the two solutions
$s_*=\pm \mu \lambda$.
We will denote with $C_i^+$ the critical point associated with $s_*=\mu \lambda$ and with $C_i^-$ the ones associated with $s_*=-\mu \lambda$ ($i=1, 2, 3, 4, 5$).
The stability regions of points $C_i^+$ and $C_i^-$ ($i=1,2,3,4,5$) for $\mu=1$ in the $(\lambda,\alpha)$ parameter space are given in Figs.~\ref{fig:c5_I_quint_region_sinh} and \ref{fig:c5_I_quint_region_sinh_neg} respectively.
Note the symmetry between the stability regions of points $C_i^+$ and points $C_i^-$.
This is due to the invariance of the hyperbolic potential under the $\lambda \mapsto -\lambda$ transformation.
The properties of the critical points for the hyperbolic potential are as follow:
\begin{itemize}

\item {\bf Points $C_1^+$, $C_1^-$}: These points correspond to decelerated, stiff matter dominated universes ($w_{\rm tot}=1$). Point $C_1^+$ is stable whenever $\alpha<-\frac{\sqrt{6}}{2}$, $\mu\lambda>\sqrt{6}$, $\lambda>0$, otherwise it is saddle. Point $C_1^-$ is stable whenever $\alpha<-\frac{\sqrt{6}}{2}$, $\mu\lambda<-\sqrt{6}$, $\lambda<0$, otherwise it is saddle.

\item {\bf Points $C_2^+$, $C_2^-$}: These points correspond to a decelerated, scaling solutions. Point $C_2^+$ is stable whenever $\alpha^2<3$, $\alpha\,\mu\lambda<-3$, $\alpha\,\lambda<0$, otherwise it is saddle. Point $C_2^-$ is stable whenever $\alpha^2<3$, $\alpha\,\mu\lambda>3$, $\alpha\,\lambda>0$, otherwise it is saddle.

\item {\bf Points $C_3^+$, $C_3^-$}: These points correspond to an accelerated solution ($q=-1$) with negative DM energy density. Point $C_3^+$ corresponds to a late time attractor (stable spiral) for $\mu >0$ and $\alpha\,\lambda>0$. Point $C_3^-$ corresponds to a late time attractor  for $\mu >0$ and $\alpha\,\lambda<0$.

\item {\bf Points $C_4^+$, $C_4^-$}: These points exist when $\mu^2\lambda^2<6$. They correspond to accelerated, scalar field dominated universes. Point $C_4^+$ is stable whenever $\mu\lambda(\alpha+\mu\lambda)<3$, $\mu<0$, otherwise it is saddle. Point $C_4^-$ is stable whenever $\mu\lambda(\alpha-\mu\lambda)>-3$, $\mu<0$, otherwise it is saddle.

\item {\bf Points $C_5^+$, $C_5^-$}: These points are cosmologically interesting as they correspond to late time accelerated scaling solutions for some values of the parameters $\alpha$, $\mu$ and $\lambda$ (cf.~Fig.~\ref{fig:stability_region_I_sinh}).
For example, if we take $\alpha=-2.8$, $\lambda=-2$, $\mu=1$, we obtain $\lambda_1=-1.25$, $\lambda_2=-1.18-3.38\,i$, $\lambda_3=-1.18+3.38\,i$, $\lambda_4=-2.5$, with $q=-0.37$, $\Omega_m=0.28$, $\Omega_{\phi}=0.71$ for point $C_5^+$, while point $C_5^-$ is saddle in nature.
On the other hand for $\alpha=-2.8$, $\lambda=2$, $\mu=1$, point $C_5^-$ is stable but point $C_5^+$ is saddle.
This implies that choosing the appropriate combination of parameters, critical points $C_5^+$ and $C_5^-$  describe a late time accelerated scaling attractor (stable spiral) with DM and DE energy density values in agreement with present observations.

\item {\bf Set of points $C_6$}: This normally hyperbolic set of critical points $C_6$ corresponds to a late time attractor only if $\mu<0$.

\end{itemize}

\begin{figure}
\centering
\subfigure[]{%
\includegraphics[width=6cm,height=4cm]{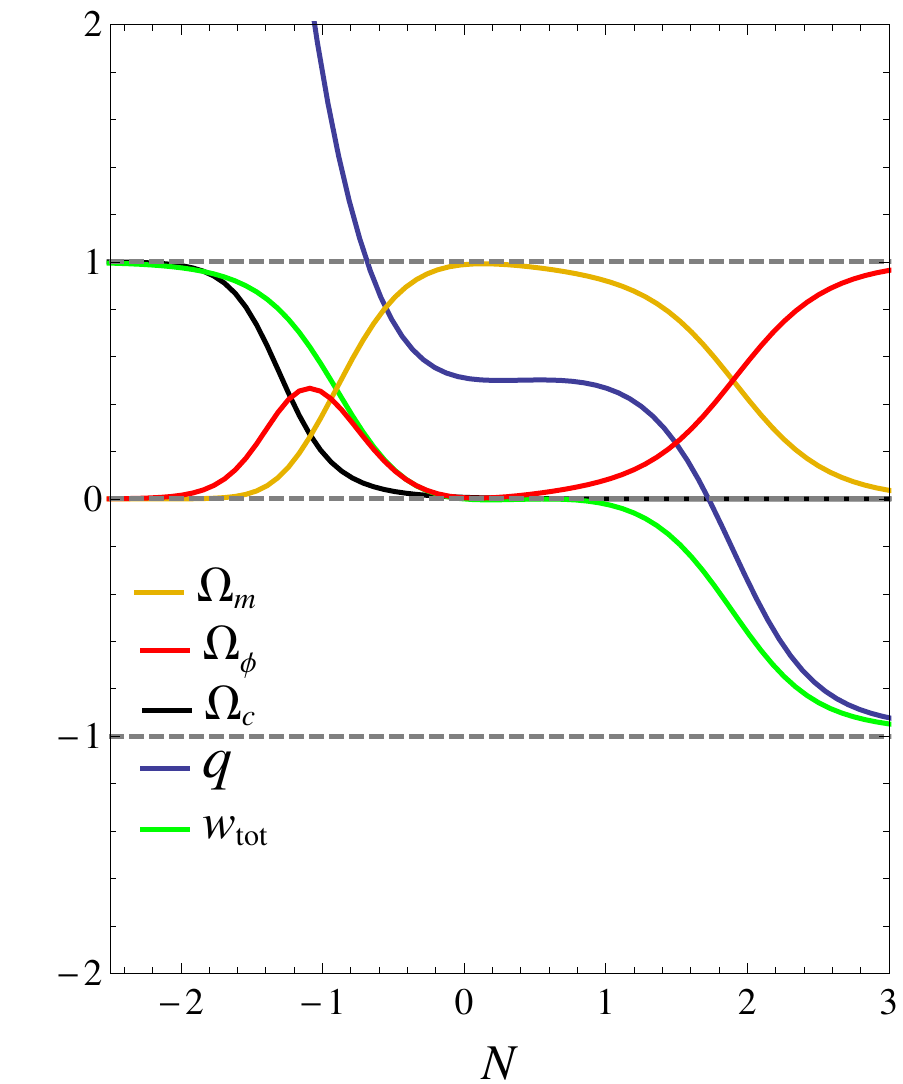}\label{fig:weff_sinh_quin_I}}
\qquad
\subfigure[]{%
\includegraphics[width=6cm,height=4cm]{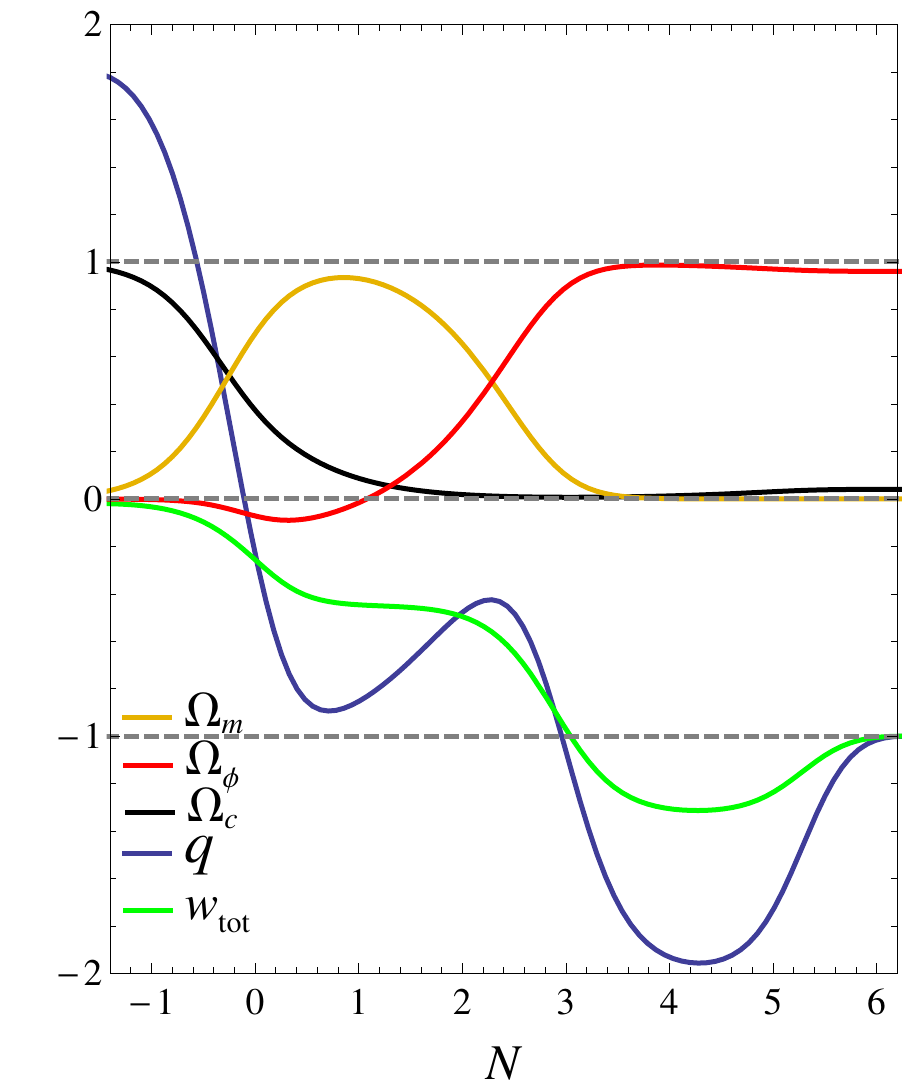}\label{fig:weff_sinh_phan_I}}
\caption{Plot of the relative matter energy density $\Omega_m$, the relative DE energy density $\Omega_\phi$, the relative energy density due to loop quantum corrections $\Omega_c$, the total EoS $w_{\rm tot}$ and the deceleration parameter $q$ versus $N$ for the potential $V(\phi)=V_0[\cosh(\lambda\phi)]^{-\mu}$. Here we assume $\lambda=0.5$, $\mu=-2$, $\alpha=0.3$ and $\epsilon=1$ in panel (a) whereas $\lambda=0.5$, $\mu=-2$, $\alpha=1$ and $\epsilon=-1$ in panel (b).
}
\label{fig:weff_sinh_I}
\end{figure}

\noindent
Fig.~\ref{fig:weff_sinh_quin_I} shows the contribution of loop quantum gravity corrections at early time together with the evolution of the relevant cosmological parameters \eqref{11}--\eqref{15}.
One particular trajectory is considered which evolves from a matter phase dominated by interacting energy (point $C_2$) and settles in the DE dominated point $C_4$.
This model can thus describe the late time DE dominated observed phase of our universe; see Sec.~\ref{sec:cosmological_implications} for further discussions.

\subsubsection*{Example 2: $V=\frac{M^{4+n}}{\phi^n}$ }

In this example we consider the inverse power-law potential $V(\phi)=\frac{M^{4+n}}{\phi^n}$ (where $M$ is a mass scale, while $n$ is a dimensionless parameter), which can lead to tracking behavior in EC \cite{Steinhardt:1999nw}.
For this potential, the function $f(s)$ is given by
 \begin{equation}\label{20}
 f(s)=\frac{s^2}{n} \,,
 \end{equation}
meaning that
\begin{equation}
 s_*=0  \hspace{1.5cm}\mbox{and}\hspace{1.5cm} df(s_*)=\frac{2s_*}{n} = 0 \,.
\end{equation}
This implies that in this case we have only one copy for each of the points $C_1$, $C_2$, $C_3$, $C_4$, $C_5$ listed in Table~\ref{Table1}.
The properties of the critical points are as follow:

\begin{figure}
\centering
\includegraphics[width=6cm,height=4cm]{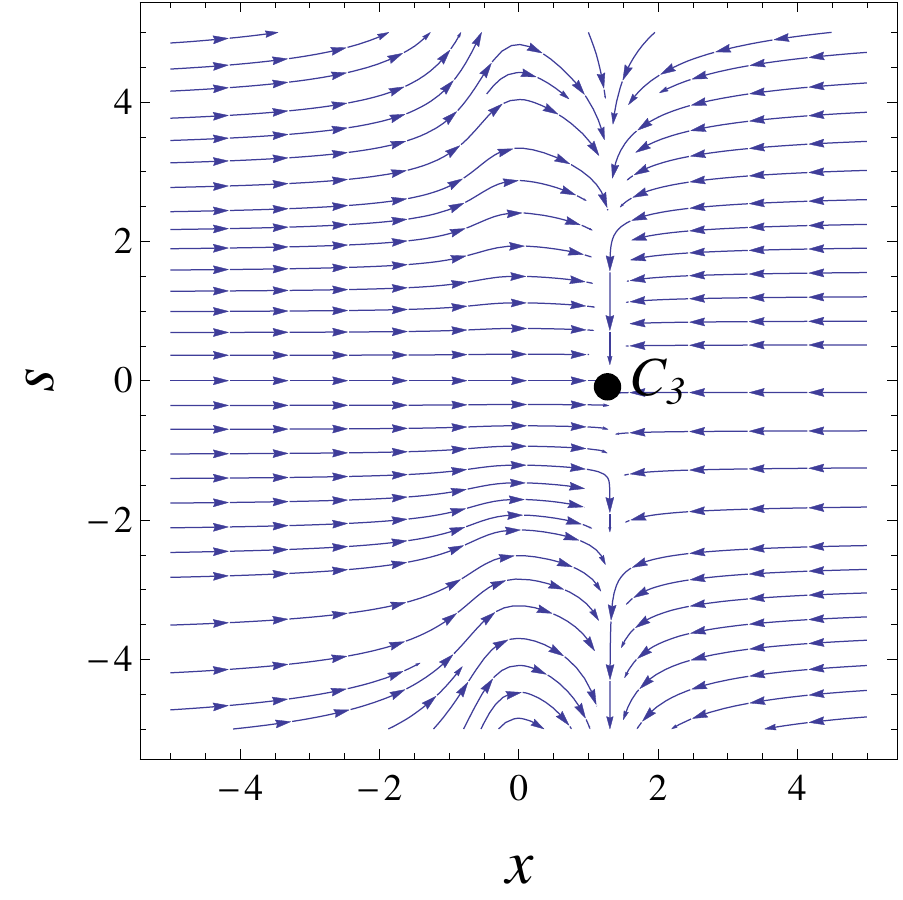}
\caption{ Stream plot projection of the dynamical flow of the system (\ref{x_I})-(\ref{s_I}) on the $(x,0,1.66,s)$ slice near the point $C_3$. Here we considered the potential $V(\phi)=\frac{M^{4+n}}{\phi^n}$ with $\alpha=1$, $n=10$ and $\epsilon=1$.}
\label{fig:strm_c3_powerlaw_quin_I}
\end{figure}
\begin{figure}
\centering
\subfigure[]{%
\includegraphics[width=6cm,height=4cm]{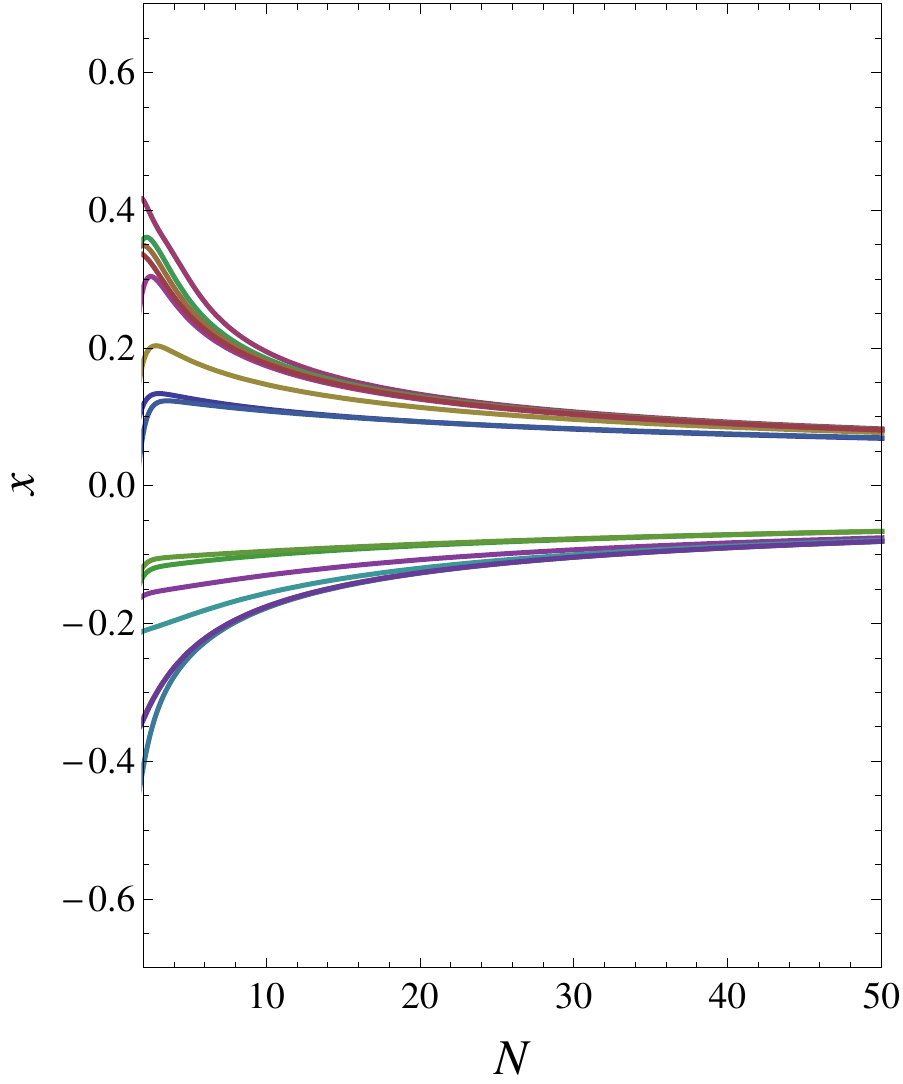}\label{fig:perturb_c4_powerlaw_x}}
\qquad
\subfigure[]{%
\includegraphics[width=6cm,height=4cm]{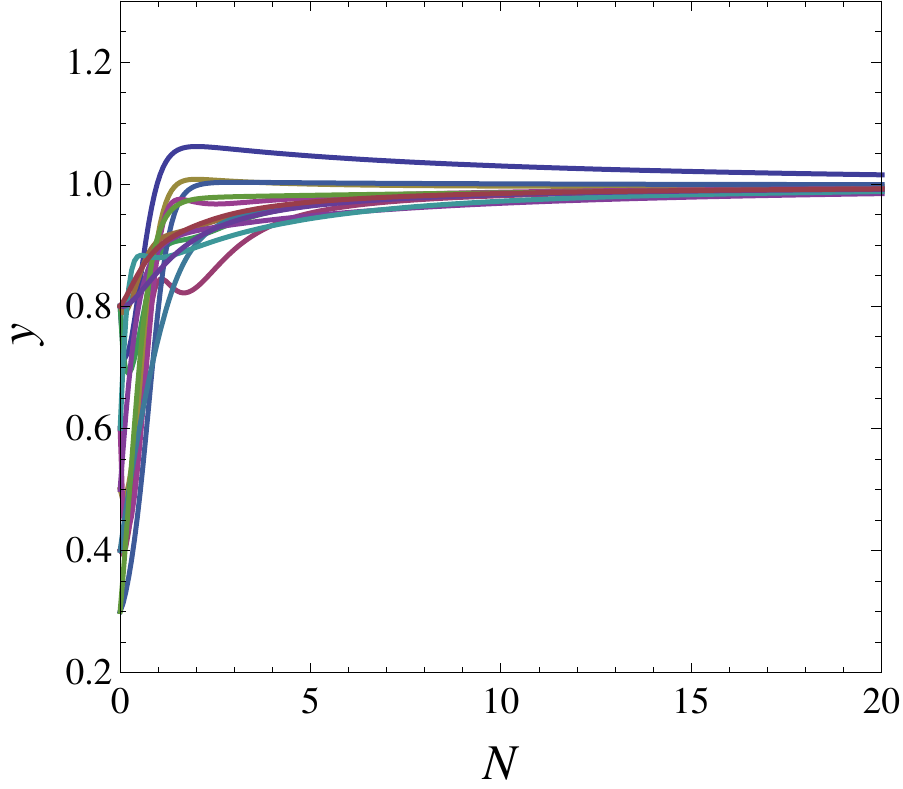}\label{fig:perturb_c4_powerlaw_y}}
\qquad
\subfigure[]{%
\includegraphics[width=6cm,height=4cm]{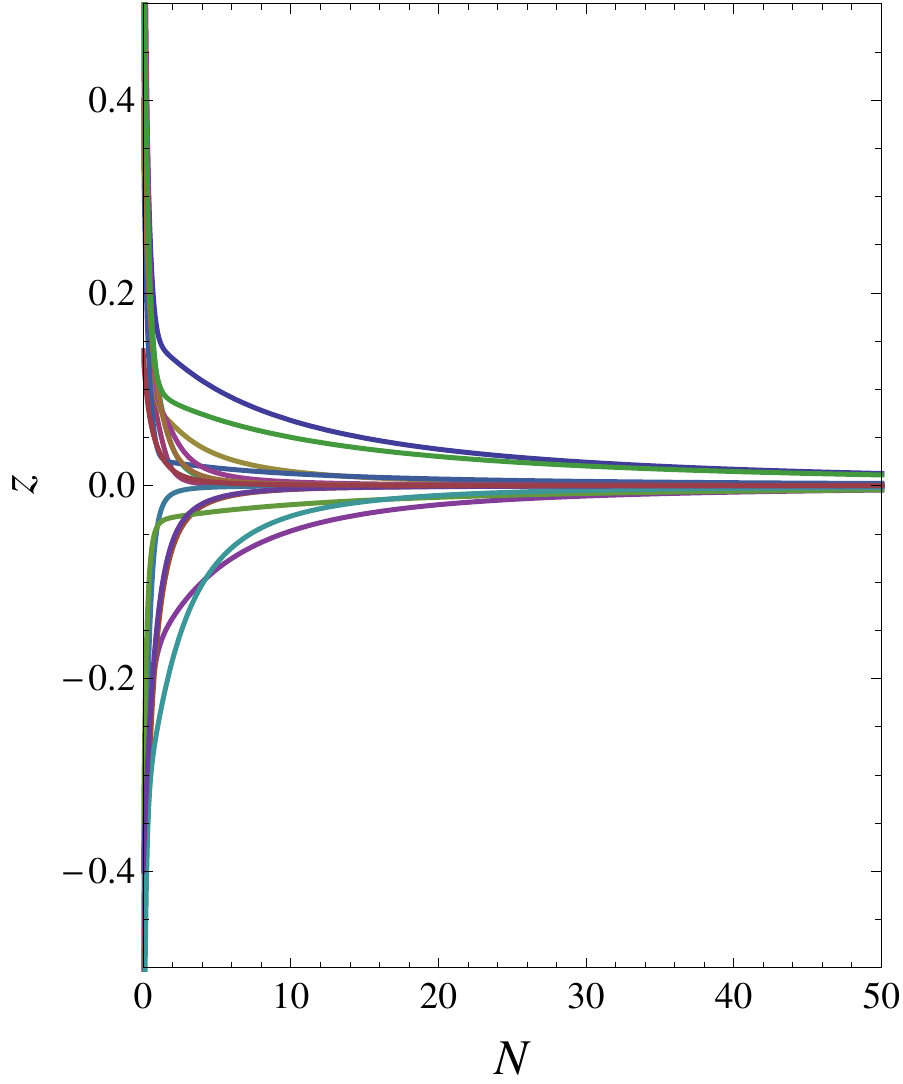}\label{fig:perturb_c4_powerlaw_z}}
\qquad
\subfigure[]{%
\includegraphics[width=6cm,height=4cm]{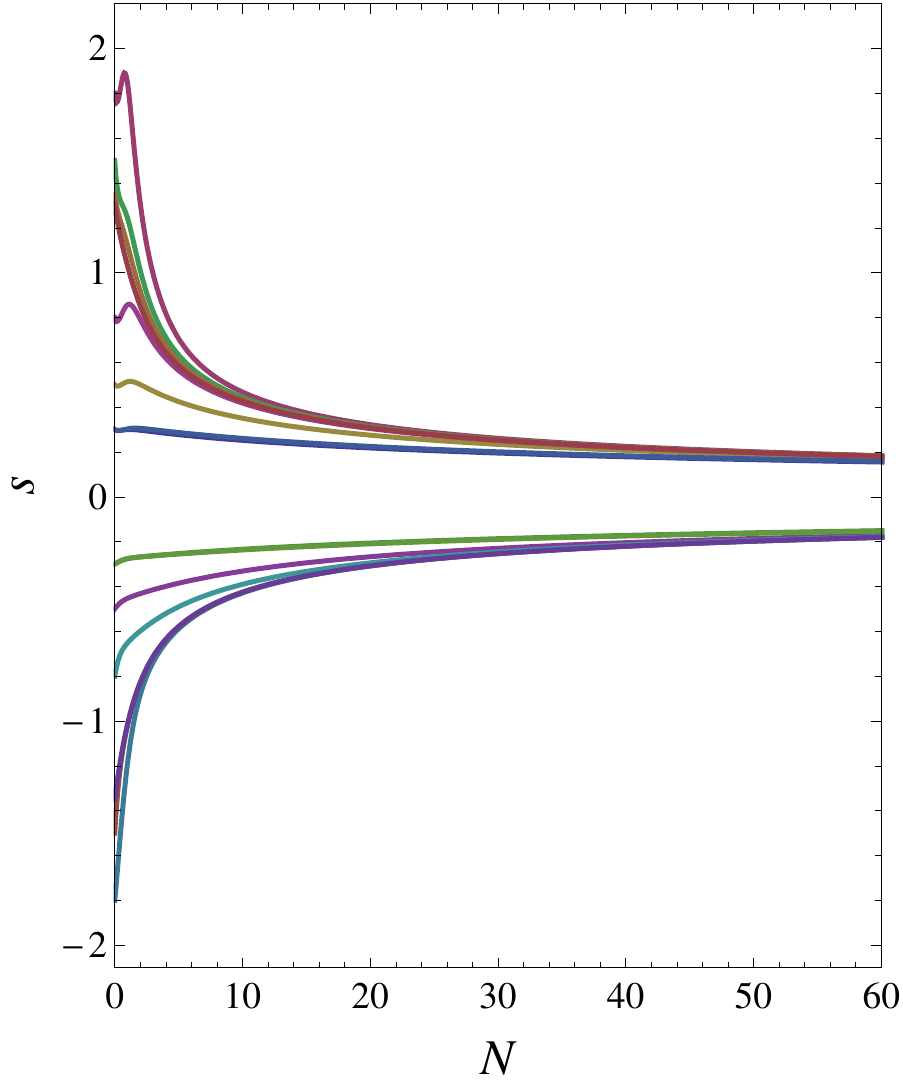}\label{fig:perturb_c4_powerlaw_s}}
\caption{Evolution of trajectories in the neighbourhood of critical point $C_4$ projected on the $x$ (a), $y$ (b), $z$ (c) and (s) axis respectively. Here we assume the potential $V(\phi)=\frac{M^{4+n}}{\phi^n}$ with $\alpha=1$, $n=4$, $\epsilon=1$.}
\label{fig:perturb_c4_powerlaw}
\end{figure}

\begin{itemize}

 \item {\bf Point $C_1$}: This describes a decelerated, stiff matter dominated point ($w_{\rm tot}=1$) and is always saddle.

 \item {\bf Point $C_2$}: This point corresponds to a decelerated, scaling solution. It is always saddle, in contrast with the hyperbolic potential for which it can be stable.

 \item {\bf Point $C_3$}: This point corresponds to an accelerated solution with negative DM energy density. It is a non-hyperbolic point, for which linear stability fails to determine its stability. In order to determine its stability we numerically stream plotted projections onto the $(x,s)$ slice of the dynamical flow around its neighbourhood, as shown for example in Fig.~\ref{fig:strm_c3_powerlaw_quin_I}.
 For several interesting combinations of the model parameters, we have checked numerically that this point cannot be stable as some trajectories are always repelled from it.

 \item {\bf Point $C_4$}: Point $C_4$ reduces to point $(0,1,0,0)$ where the scalar field potential energy dominates.
 It is a non-hyperbolic point, so linear stability fails to determine its stability.
 Usually the stability of this type of critical point can be determined analytically by employing center manifold theory, otherwise one can also use numerical methods; see e.g.~\cite{Dutta:2016dnt,Dutta:2016bbs,Roy:2014hsa} for some applications in the recent literature.
 Numerically one can plot the projection of trajectories around the critical point separately on the $x$, $y$, $z$ and $s$ axes (see Fig.~\ref{fig:perturb_c4_powerlaw} for example).
 We have checked numerically that nearby trajectories asymptotically approach the coordinates corresponding to point $C_4$ only for $n>0$, while the parameter $\alpha$ can be arbitrary.
 Hence depending on the choices of the model parameters this point can correspond to a late time attractor.
 
 \item {\bf Point $C_5$}: This point corresponds to an accelerated solution ($q=-1$) with negative DM energy density $\Omega_{m}<0$. It is always a saddle since the eigenvalues $\lambda_2$ and $\lambda_3$ are always of opposite sign.

 \item {\bf Set of points $C_6$}: For the inverse power-law potential, this set of non-isolated critical points is non-hyperbolic but not normally hyperbolic since it presents two vanishing eigenvalues.
 We use numerical techniques to check stability by plotting the projection of nearby trajectories separately on the $x$, $y$, $z$ and $s$ axes.
 We find that these trajectories approach the set of points $C_6$ as $N\rightarrow \infty$ for $n>0$.
 In fact the effective EoS parameter is always $-1$  for any point belonging to this set.
 \end{itemize}

\subsection{\bf Phantom dark energy $(\epsilon=-1)$}

We now turn our attention to phantom DE with $\epsilon=-1$. Region of stability in the $(s_*,\alpha)$ parameter space for the critical points listed in Table~\ref{Table1} are given in Fig.~\ref{fig:c5_I_phan_region}.
Their properties are as follow:
\begin{itemize}

 \item {\bf Point $C_1$}: Point $C_1$ does not exist for $\epsilon=-1$.

 \item {\bf Point $C_2$}: This point corresponds to an accelerated solution for $\alpha^2>\frac{1}{2}$, with negative DE energy density parameter. It is stable when $\alpha^2<\frac{3}{2}$, $2\alpha\,(\alpha+s_*)>3$ and $\alpha\,df(s_*)<0$.
 
 \item {\bf Point $C_3$ }: This point corresponds to an accelerated solution with negative DE energy density parameter. It is a stable node if $\frac{3}{2}<\alpha^2<2$, $\alpha\,s_*>0$, $\alpha\,df(s_*)>0$, it is stable spiral if $\alpha^2>2$, $\alpha\,s_*>0$, $\alpha\,df(s_*)>0$, otherwise it is saddle.
 
 \item {\bf Point $C_4$}: Point $C_4$ exists for any values $s_*$. It corresponds to an accelerated scalar field dominated universe and it is saddle. Although in LQC this point cannot be a late time attractor, it is found to be stable in EC where it corresponds to a future big rip singularity \cite{Guo:2004xx}.
 This behavior was first noticed in the case of an exponential potential \cite{Fu:2008gh} and constitutes an interesting example of how cosmological singularities can be avoided by LQG effects.
 
 \item {\bf Point $C_5$}: Point $C_5$ corresponds to an accelerated, scaling solution for some values of parameter $\alpha$ and $s_*$. However this point is always saddle as the eigenvalues $\lambda_2$ and $\lambda_3$ have opposite sign.

 \item {\bf Set of points $C_6$}: As in the case of the quintessence field, this non-isolated set of critical points $C_6$ is a normally hyperbolic set.
 It corresponds to a late time attractor only if $f(0)<0$.
 It is a stable node if $-\frac{3}{4}(1-z)<f(0)<0$, it is stable spiral if $f(0)<-\frac{3}{4}(1-z)$, otherwise it is saddle.
 Further investigation is required when $f(0)=0$ and we shall postpone this once a particular scalar field potential has been chosen. Again this set of points is interesting as it shows the effects of loop quantum corrections in determining the late time accelerated phase of the universe.
 Interestingly this point is a late time attractor which is not present in the case of the exponential potential \cite{Fu:2008gh}.
 \end{itemize}

\noindent
From the above general analysis, we note that from Fig.~\ref{fig:c5_I_phan_region} points $C_2$, $C_3$ cannot be late time attractors simultaneously.
Depending on the choice of the scalar field potential and the initial conditions, we thus find that the universe evolves towards an accelerated, scalar field dominated set of critical points $C_6$ or towards a negative DE density critical points $C_2$, $C_3$.
Again since the existence and stability properties of the critical points depend heavily on $s_*$, $df(s_*)$ and $f(0)$, in what follows we analyze their properties selecting two specific potentials.

\subsubsection*{Example 1: $V=V_0\,\cosh^{-\mu}(\lambda\phi)$}

\begin{figure}
\centering
\subfigure[]{%
\includegraphics[width=6cm,height=4cm]{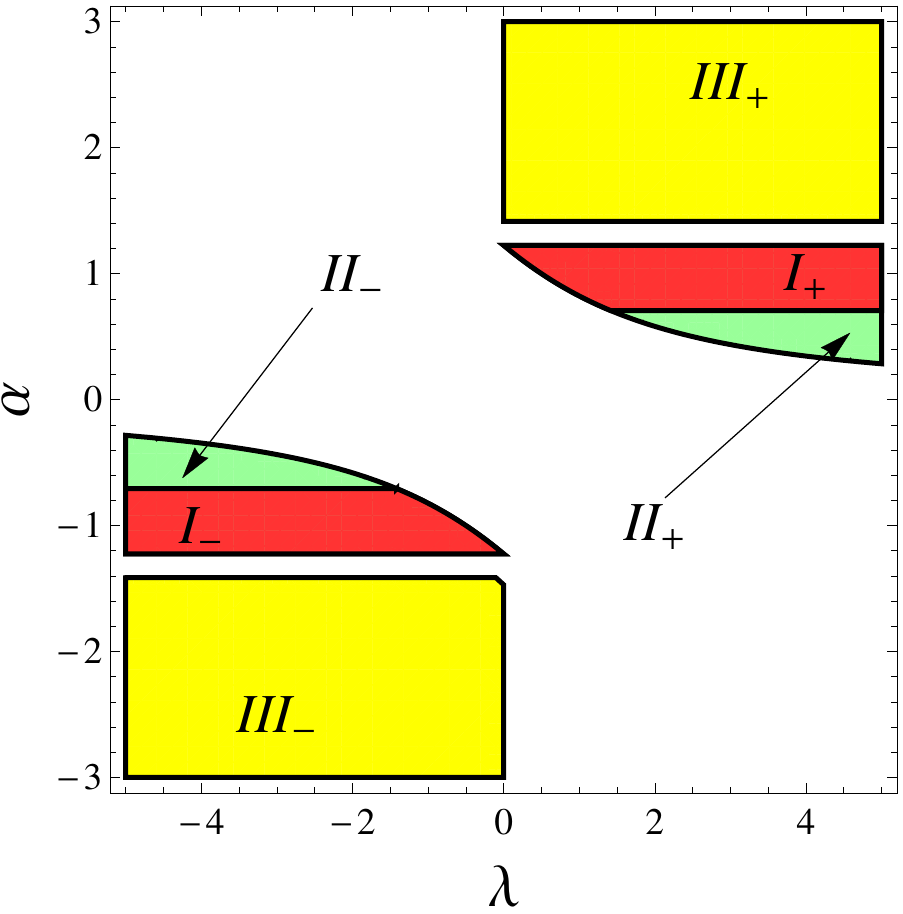}\label{fig:c5_I_phan_region_sinh}}
\qquad
\subfigure[]{%
\includegraphics[width=6cm,height=4cm]{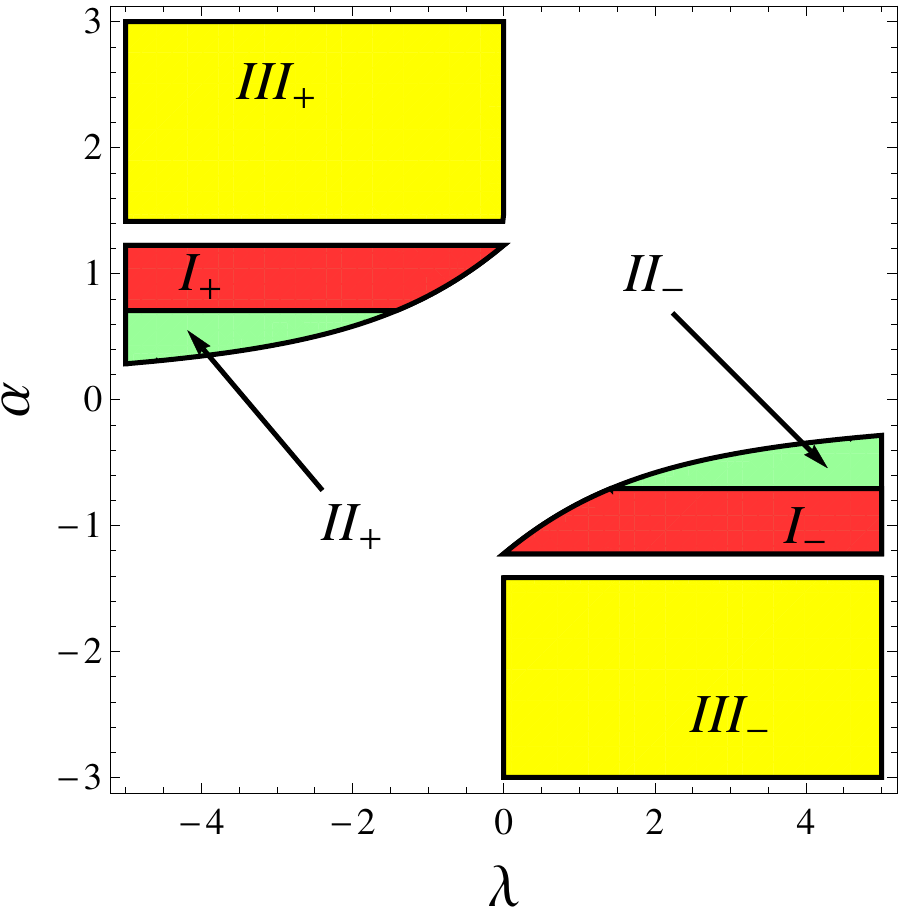}\label{fig:c5_I_phan_region_sinh_s_negative}}
\caption{(a). Stability regions of points $C_i^+$ ($i=2, 3$) in $(\lambda,\alpha)$ parameter space (b). Stability regions of points $C_i^-$ ($i=2, 3$) in $(\lambda,\alpha)$ parameter space with potential $V=V_0\,\cosh^{-\mu}(\lambda\phi)$ by taking $\mu=1$. In respective panel, Regions $I_+$, $II_+$, $I_-$, $II_-$ represent regions of stability of point $C_2^+$ ($C_2^-$). Region $III_+$, $III_-$ represent regions of stability of point $C_3^+$ ($C_3^-$). Regions $I_+$, $I_-$ represents regions of acceleration of point $C_2^+$ ($C_2^-$). Here $\epsilon=-1$.}
\label{fig:stability_region_I_sinh_phan}
\end{figure}

Here we report the properties of the critical points in Table~\ref{Table1} for $\epsilon=-1$ and the potential $V=V_0\,\cosh^{-\mu}(\lambda\phi)$.
Again here we have two copies of each critical points $C_2$, $C_3$, $C_4$, $C_5$, one for each of the two solutions $s_*=\pm \mu \lambda$.
As in the case of quintessence each point $C_i$ with $s_*=\mu \lambda$ is denoted as $C_i^+$ while point with $s_*=-\mu \lambda$ are denoted as $C_i^-$ ($i=2,3,4,5$).
The stability regions of $C_2^+$, $C_3^+$ are given in Fig.~\ref{fig:c5_I_phan_region_sinh} and that of $C_2^-$, $C_3^-$ are given in Fig.~\ref{fig:c5_I_phan_region_sinh_s_negative}.
Again the symmetry of these two figures is due to the invariance of the hyperbolic potential under the $\lambda \mapsto -\lambda$ transformation.

\begin{itemize}

 \item {\bf Points $C_2^+$, $C_2^-$}: These points correspond to an accelerated solution for $\alpha^2>\frac{1}{2}$, with negative DE energy density parameter. Point $C_2^+$ is stable when $\alpha^2<\frac{3}{2}$, $2\alpha\,(\alpha+\lambda\mu)>3$ and $\alpha\,\lambda<0$. Point $C_2^-$ is stable when $\alpha^2<\frac{3}{2}$, $2\alpha\,(\alpha-\lambda\mu)>3$ and $\alpha\,\lambda>0$.

 \item {\bf Points $C_3^+$, $C_3^-$}: Point $C_3^+$ is a stable node if $\frac{3}{2}<\alpha^2<2$, $\mu>0$, $\alpha\,\lambda>0$, it is a stable spiral if $\alpha^2>2$, $\mu>0$, $\alpha\,\lambda>0$, otherwise it is saddle. Point $C_3^-$ is a stable node if $\frac{3}{2}<\alpha^2<2$, $\mu>0$, $\alpha\,\lambda<0$, it is a stable spiral if $\alpha^2>2$, $\mu>0$, $\alpha\,\lambda<0$, otherwise it is saddle.

 \item {\bf Points $C_4^+$, $C_4^-$}: They correspond to an accelerated, scalar field dominated universe. However both of them are saddle in nature for any choice of model parameters.

 \item {\bf Points $C_5^+$, $C_5^-$}: These accelerated scaling solutions are saddle in nature for any choice of model parameters.

 \item {\bf Set of points $C_6$}: This non-isolated normally hyperbolic set of critical points $C_6$ corresponds to a late time attractor only if $\mu>0$.

\end{itemize}

\noindent
From the analysis above, we see that depending on the initial conditions the universe can evolve towards an accelerated, scalar field dominated set of critical points $C_6$.
This can be seen in Fig.~\ref{fig:weff_sinh_phan_I} which shows the evolution of cosmological parameters towards an accelerating phase $q=-1$ (point $C_6$).
Another interesting observation is that a transient phantom epoch described by points $C_4^+$, $C_4^-$ ($q=-1.5$) might be possible.
Although the universe undergoes a super-accelerated expansion ($\dot{H}>0$) at this point, the future big rip singularity is avoided by loop quantum effects which turn this point from an attractor to a saddle.

\subsubsection{Example 2: $V=\frac{M^{4+n}}{\phi^n}$ }

We recall that for this potential we have $s_*=0$ and $df(s_*)=0$, implying in particular that there is only one copy for each of the point $C_2$, $C_3$, $C_4$, $C_5$.
The properties of the critical points are as follow:

\begin{figure}
\centering
\subfigure[]{%
\includegraphics[width=6cm,height=4cm]{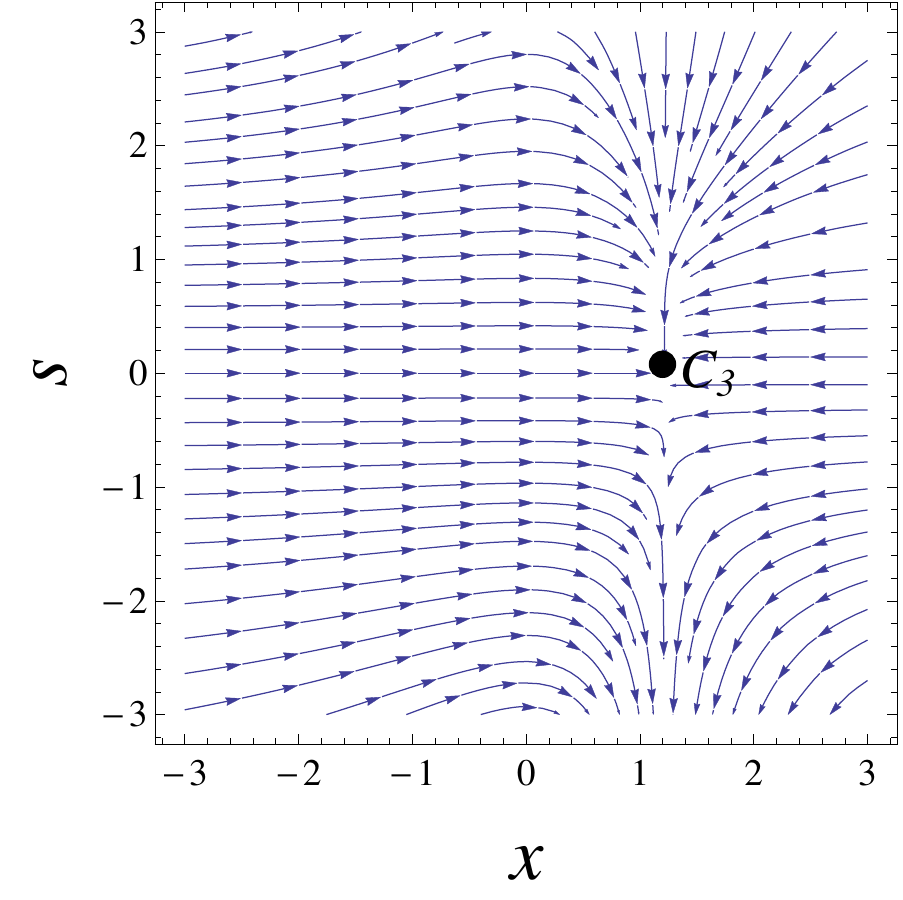}\label{fig:strm_c3_powerlaw_phan_I}}
\qquad
\subfigure[]{%
\includegraphics[width=6cm,height=4cm]{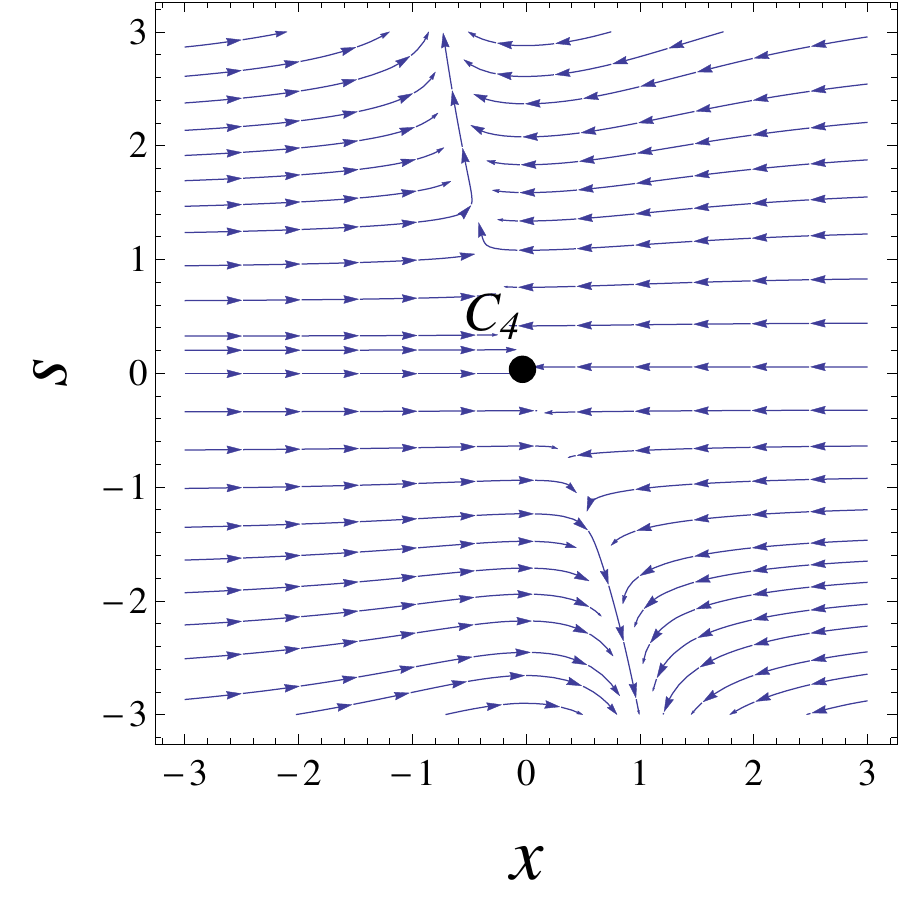}\label{fig:strm_c4_powerlaw_phan_I}}
\caption{(a). Stream plot projection of the dynamical flow of the system (\ref{x_I})-(\ref{s_I}) on the $(x,0,0.33,s)$ slice near the point $C_3$ (b). Stream plot projection of the dynamical flow of the system (\ref{x_I})-(\ref{s_I}) on the $(x,1,0,s)$ slice near the point $C_4$. Here we consider the potential $V(\phi)=\frac{M^{4+n}}{\phi^n}$ with $\alpha=1$, $n=10$, $\epsilon=-1$.}
\label{fig:strm_c3_c4_powerlaw_phan_I}
\end{figure}

\begin{itemize}
\item {\bf Point $C_2$}: Point $C_2$ describes a solution with negative DE energy density. It is always saddle.

\item {\bf Point $C_3$}: Point $C_3$ corresponds to an accelerated solution with negative DE density. We have numerically checked that it is never stable for any choice of parameters. An example of the flow around this point projected on the $(x,s)$-plane is given in Fig.~\ref{fig:strm_c3_powerlaw_phan_I}.

\item {\bf Point $C_4$}: Point $C_4$ exists for any values $\alpha$, $\mu$ and $\lambda$. It is a non-hyperbolic critical point and it is not stable as confirmed numerically. An example of the dynamical flow around this point, projected on the $(x,s)$-plane, is given in Fig.~\ref{fig:strm_c4_powerlaw_phan_I} for some values of the model parameters.

\item {\bf Point $C_5$}: Point $C_5$ corresponds to an accelerated, scaling solution. This point is always saddle.

\item {\bf Set of points $C_6$}: Since this set contains point $C_4$ and trajectories on the $(x,s)$ sub-space do not approach the coordinates of point $C_4$, this set cannot be a late time attractor, as it is for the hyperbolic potential. This has been checked numerically for different set of model parameters of phenomenological interest.

\end{itemize}
From this analysis, we see that there is no finite late time attractor.
This interacting phantom model cannot thus explain the late time behavior of our universe, although future big rip singularities appearing in EC can be avoided.


\section{Interacting model II: $\bf Q=\beta\dot\rho_\phi$}
\label{section4}

The dynamical investigation of this interaction term for quintessence dark energy in the framework of standard EC has been recently
studied in \cite{Shahalam:2015sja} where it was found that this model can alleviate the coincidence problem.
A generalised form of this interaction has been used in \cite{Wei:2010fz,Wei:2010cs} to build a coupled DE model where the sign of the dark interaction changes during the cosmological history.
The specific interaction $Q=\beta\dot\rho_\phi$ is also motivated from the dimensional point of view since $Q$ has the dimension of a time rate of energy density \cite{Shahalam:2015sja}.
Again in this section we will only expose the dynamical system analysis of the interacting model under study, while leaving the discussion on the cosmological implications to section~\ref{sec:cosmological_implications}.

For this interaction the system (\ref{x_prime})-(\ref{s_prime}) becomes
\begin{align}
x'&= -\frac{3x}{(1+\beta)}+\frac{1}{\epsilon}~\sqrt{\frac{3}{2}}~sy^2  +x\Big[\frac{3}{2} \Big(\frac{1}{1-z}-\epsilon x^2-y^2\Big)+\epsilon3x^2\Big](1-2z),\label{x_II}\\
y'&=-\sqrt{\frac{3}{2}}~sxy +y\Big[\frac{3}{2}\Big(\frac{1}{1-z}-\epsilon x^2-y^2\Big)+\epsilon3x^2\Big](1-2z),\label{y_II}\\
z'&=-3z-3z(1-z)(\epsilon x^2-y^2),\label{z_II}\\
s'&=-\sqrt{6}~xf(s)\label{s_II}\,.
\end{align}
It can be seen that the system (\ref{x_II})-(\ref{s_II}) is symmetric under the transformation $y \rightarrow -y$.
It is also
symmetric under the transformation $(x,s) \rightarrow (-x,-s)$ for
potential with $f(s)=f(-s)$. The critical points of the system
(\ref{x_II})-(\ref{s_II}) and their properties are shown in
Table~\ref{Table3} and the corresponding eigenvalues of the
Jacobian matrix are given in Table~\ref{Table4}.
The dynamics of this model is simple if compared to that of the interacting model I analysed in section~\ref{section3}.
The system (\ref{x_II})-(\ref{s_II}) contains only two critical points and
two non-isolated set of critical points.
The non-isolated set $D_1$
is completely independent from the scalar field for its existence
and stability. On the other hand, critical points $D_2$, $D_3$ and the
non-isolated set $D_4$ depend on the specific choice of the scalar
field potential for their existence and stability. As before, in
what follows, we analyze the cosmological dynamics separately
for quintessence and phantom DE.

\subsection{\bf Quintessence dark energy $(\epsilon=1)$}

\begin{table}[b]
\centering

Here: $\Delta=6\, \left( \beta+1 \right) {s_*}^{2} \left( {s_*}^{2} \left( \beta+1
 \right) -6\,\epsilon \, \left( 1+2\,\beta \right)  \right) +54\,{
\epsilon }^{2}$
\begin{adjustbox}{width=1\textwidth}

\small

\begin{tabular}{|c|c|c|c|c|c|c|c|c|c|}
\hline
Point&$x$~&$y$~&$ z$~&$s$~&${\Omega}_{\phi}~$&${\Omega}_{m}~$&$q$~&$w_{\rm tot}$~&Existence\\\hline
\hline
$D_1$&0~&0~&0~&$s$~&$0$~&$1$~&$\frac{1}{2}$~&$0$~&Always\\[2ex]
$D_2$&$\sqrt{\frac{1-\beta}{\epsilon(1+\beta)}}$~&0~&$0$~&$s_{*}$~&$\frac{1-\beta}{\beta+1}$~&$\frac{2\beta}{\beta+1}$~&$\frac{2-\beta}{\beta+1}$~&$\frac{1-\beta}{1+\beta}$&$0 \leq \beta \leq 1\,(\epsilon=1)$\\&&&&&&&&&$\beta<-1$ or $\beta>1$ ($\epsilon=-1$)\\[2ex]
$D_3$&$\frac{1}{12}\frac {\sqrt {6} \left( {s_*}^{2} \left( \beta+1 \right) +3\,
\epsilon  \right) +\Delta}{s_*\epsilon \, \left( \beta+1 \right)}$~&$\frac{1}{12}\sqrt {-{\frac {2\,\Delta \left( \Delta+\sqrt {6} \left( {s_*}^{2}
 \left( \beta+1 \right) -3\,\epsilon  \right)  \right) +72\,\beta\,{
s_*}^{2}\epsilon \, \left( \beta+1 \right) -216\,{\epsilon }^{2}}{{s_*}^{
2}\epsilon \, \left( \beta+1 \right) ^{2}}}}$& ~0~ & $s_*$~&--~&--~&--~&--~&Fig. \ref{fig:d3_regplot}\\[2ex]
$D_4$&0~&$\frac{1}{\sqrt{1-z}}$~&$z$~&0~&$\frac{1}{1-z}$~&0~&$-1$~&$-1$&$z<1$\\\hline
\end{tabular}

\end{adjustbox}
\caption{Critical points of the system (\ref{x_II})-(\ref{s_II}) and values of the relevant parameters for a generic scalar field potential. ${\Omega}_{\phi}~$, ${\Omega}_{m}~$, $q$ and $w_{\rm tot}$ for $D_3$ are not shown due to their lengthy expressions.
}
\label{Table3}
\end{table}

\begin{table}
\centering
\small
\begin{tabular}{|c|c|c|c|c|c|c|}
\hline
Point & $\lambda_1$~ & $\lambda_2$~ & $\lambda_3$~ & $\lambda_4$&Stability \\\hline
\hline
&&&&&\\
$D_1$&$0$~&$-3$~&$\frac{3}{2}$~&$-\frac{3}{2}\frac{1-\beta}{\beta+1}$~&Saddle ($\epsilon=\pm 1$)\\[3ex]
$D_2$&$-\frac{6}{\beta+1}$~&$3\frac{1-\beta}{\beta+1}$~&$-\frac{\sqrt{6}s_*\Big(\sqrt{\frac{1-\beta}{\epsilon(1+\beta)}}(1+\beta)-6\Big)}{2(1+\beta)}$~&$-\sqrt{6}\sqrt{\frac{1-\beta}{\epsilon(1+\beta)}}\,df(s_*)$~&Saddle ($\epsilon=1$)\\
&&&&&$s_*\sqrt{\frac{\beta-1}{\beta+1}}>\frac{6 s_*}{\beta+1}$ ($\epsilon=-1$)\\[3ex]
$D_3$& - & - & - & - & Saddle ($\epsilon=\pm 1$)\\[3ex]
$D_4$&$0$~&$-3$~&$-\frac{3}{2(\beta+1)}\left[1+\sqrt{1-\frac{4f(0)(1+\beta)^2}{3\epsilon(1-z)}}\,\right]$~&$-\frac{3}{2(\beta+1)}\left[1-\sqrt{1-\frac{4f(0)(1+\beta)^2}{3\epsilon(1-z)}}\,\right]$~&$\beta>-1$, $\epsilon\,f(0)>0$\\\hline
\end{tabular}
\vspace{0.1cm}\\
\caption{Eigenvalues of the linearised matrix of the system (\ref{x_II})-(\ref{s_II}) evaluated at the critical points. The eigenvalues of $D_3$ are not reported due to their lengthy expressions.}
\label{Table4}
\end{table}

\begin{figure}
\centering
\subfigure[]{%
\includegraphics[width=6cm,height=4cm]{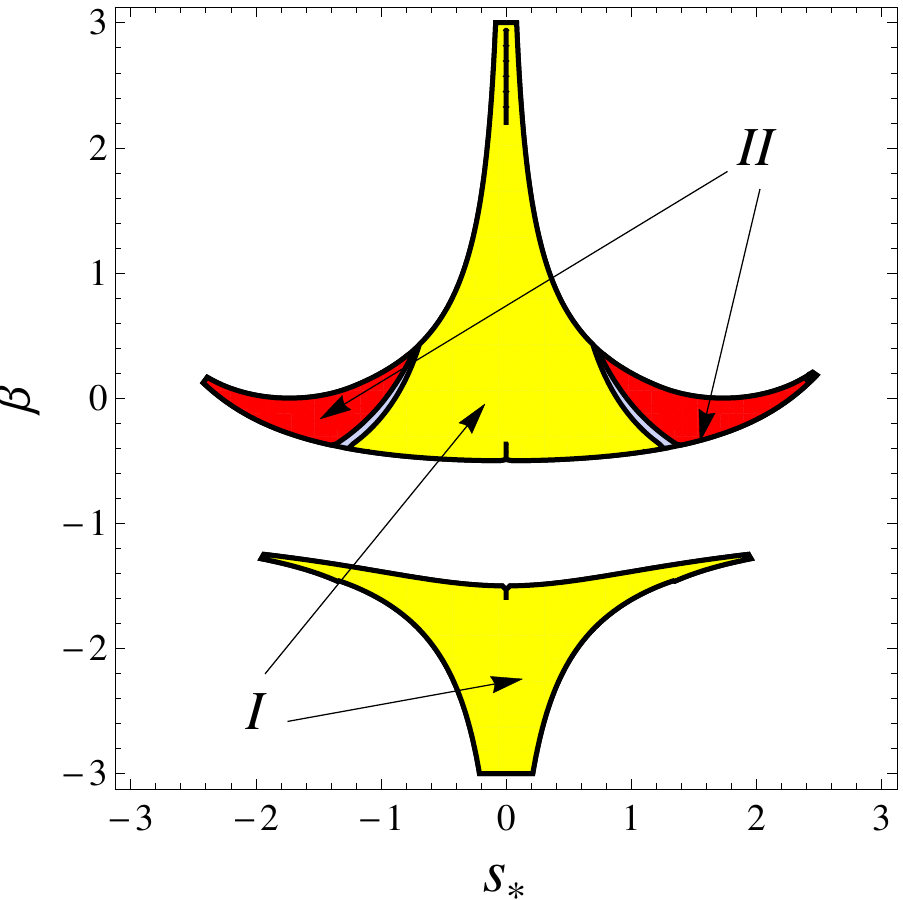}\label{fig:d3_regplot_quin}}
\qquad
\subfigure[]{%
\includegraphics[width=6cm,height=4cm]{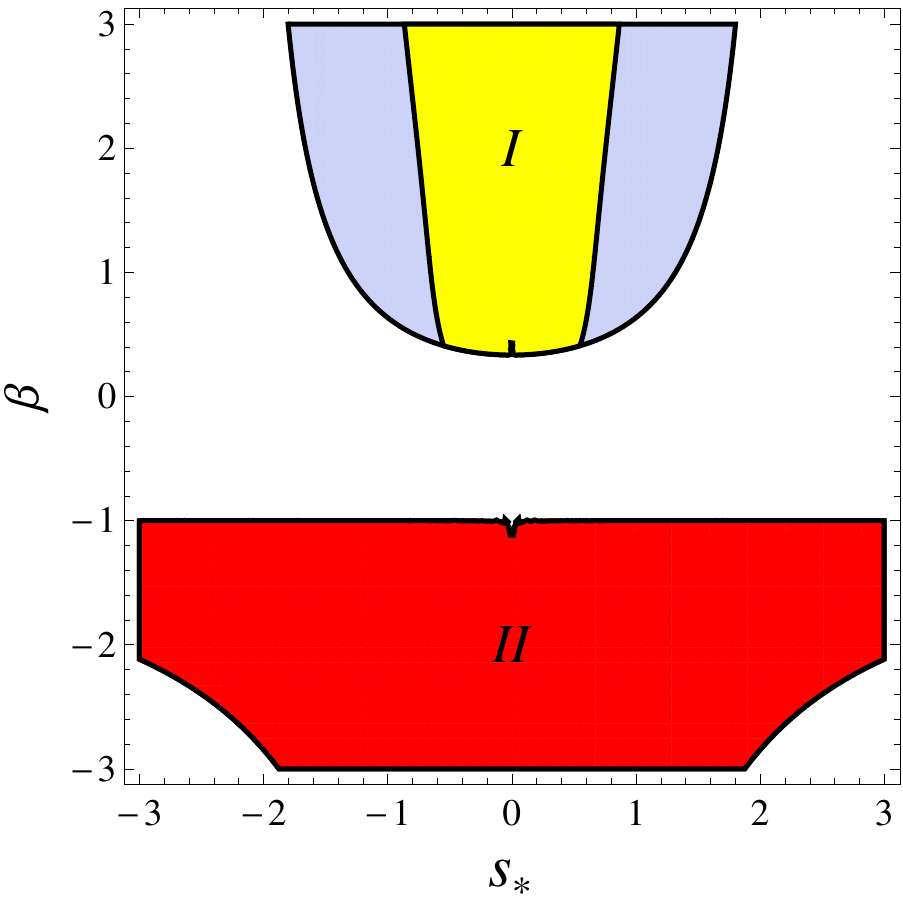}\label{fig:d3_regplot_phan}}
\caption{For both panels the whole shaded region is the region of existence of point $D_3$, region I is the region in the $(s_*,\beta)$ parameter space where the eigenvalue $\lambda_1$ of point $D_3$ is negative and region II where the eigenvalue $\lambda_3$ of point $D_3$ is negative. In panel (a) we take $\epsilon=1$ (quintessence), in panel (b) we take $\epsilon=-1$ (phantom field).
}
\label{fig:d3_regplot}
\end{figure}

The properties of the critical points in table~\ref{Table4} with $\epsilon = 1$ are the following:
\begin{itemize}

\item {\bf Set $D_1$}:  This non-isolated set of critical points corresponds to a decelerated ($q=\frac{1}{2}$), matter dominated universe ($\Omega_m=1$). It is always saddle.

\item {\bf Point $D_2$}: This point corresponds to a decelerated,
scaling solution. It is always saddle since the eigenvalues
$\lambda_1$ and $\lambda_2$ are of opposite signs within the
region of existence.

\item {\bf Point $D_3$}: The stability of this scaling solution is challenging to determine analytically due to the complicated expression of the eigenvalues of the Jacobian matrix of the system evaluated at this point.
Instead we analyzed its stability numerically by plotting the regions where $\lambda_1<0$ and $\lambda_3<0$ on the $(s_*,\beta)$ parameter space as shown in Fig.~\ref{fig:d3_regplot_quin}.
Looking at Fig.~\ref{fig:d3_regplot_quin} we can conclude that this point is always saddle since the region of negativity of the eigenvalues $\lambda_1$ and $\lambda_3$ are disjoint.
This is different from the result obtained in standard EC, where this point is a late time accelerated attractor \cite{Shahalam:2015sja}.

\item {\bf Set of points $D_4$}: The set of critical points $D_4$ exists only for $s=0$.
It is also a normally hyperbolic set with one vanishing eigenvalue if $f(0) \neq 0$.
This set corresponds to a late time attractor only if $\beta>-1$ and $f(0)>0$.
It is a stable node if $\beta>-1$ and $0<f(0)<\frac{1-z}{4(\beta+1)^2}$, it is stable spiral if $\beta>-1$ and $f(0)>\frac{1-z}{4(\beta+1)^2}$, otherwise it is saddle.
Further investigation is required when $f(0)=0$ for which we shall postpone the analysis once a particular potential has been chosen.
Again this set of points is interesting as it shows the effect of loop quantum corrections to explain the late time accelerated universe.
\end{itemize}

\noindent
From this analysis, we see that the universe can evolve from a matter dominated phase (point $D_1$) towards an accelerated scalar field dominated attracting set (set $D_4$) through a scaling solution (point $D_2$ or $D_3$).
This scenario might be used to explain the late time transition of our universe from a matter dominated phase towards a dark energy dominated phase.
Since the existence and stability behavior of points $D_2$, $D_3$ and $D_4$ depends on $s_*$, $df(s_*)$ and $f(0)$, we shall better analyze their properties for particular scalar field potentials.
In what follows we investigate again the two potentials considered in the examples above.

 \begin{figure}
\centering
\subfigure[]{%
\includegraphics[width=6cm,height=4cm]{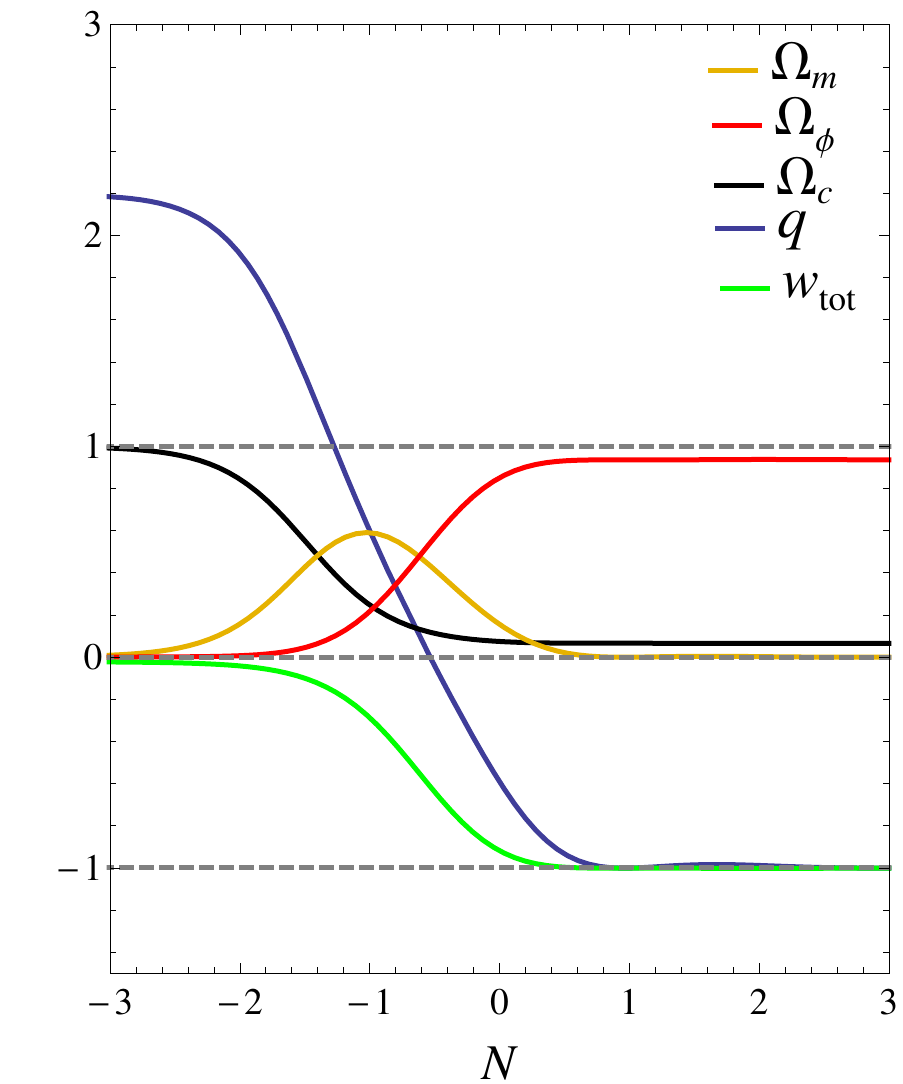}\label{fig:weff_sinh_quin_II}}
\qquad
\subfigure[]{%
\includegraphics[width=6cm,height=4cm]{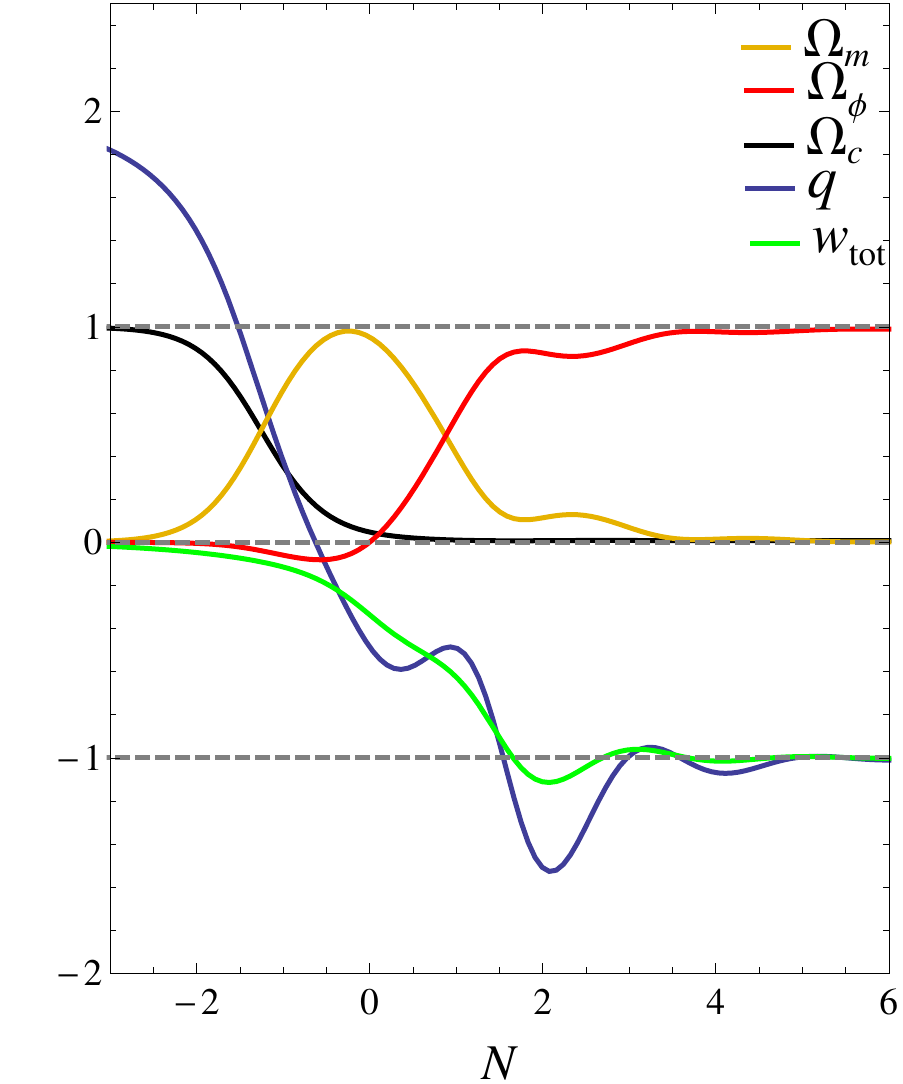}\label{fig:weff_sinh_phan_II}}
\caption{Plot of $q$ versus $N$ with the potential $V(\phi)=V_0[\cosh(\lambda\phi)]^{-\mu}$. Here we take $\lambda=1$, $\mu=-1$, $\beta=1$ and $\epsilon=1$ in panel (a) whereas $\lambda=1$, $\mu=1$, $\beta=2$ and $\epsilon=-1$  in panel(b).
}
\label{fig:weff_sinh_II}
\end{figure}

\begin{figure}
\centering
\subfigure[]{%
\includegraphics[width=6cm,height=4cm]{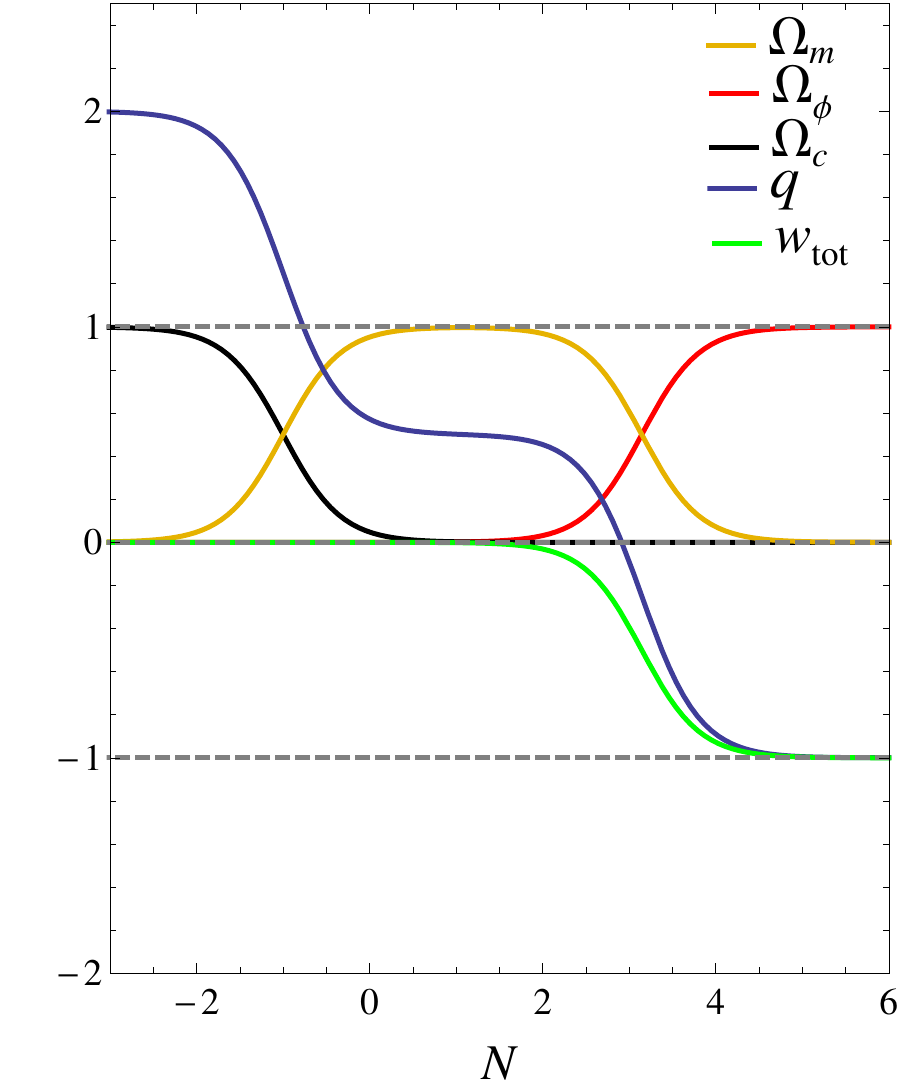}\label{fig:dm_de_quin_II}}
\qquad
\subfigure[]{%
\includegraphics[width=6cm,height=4cm]{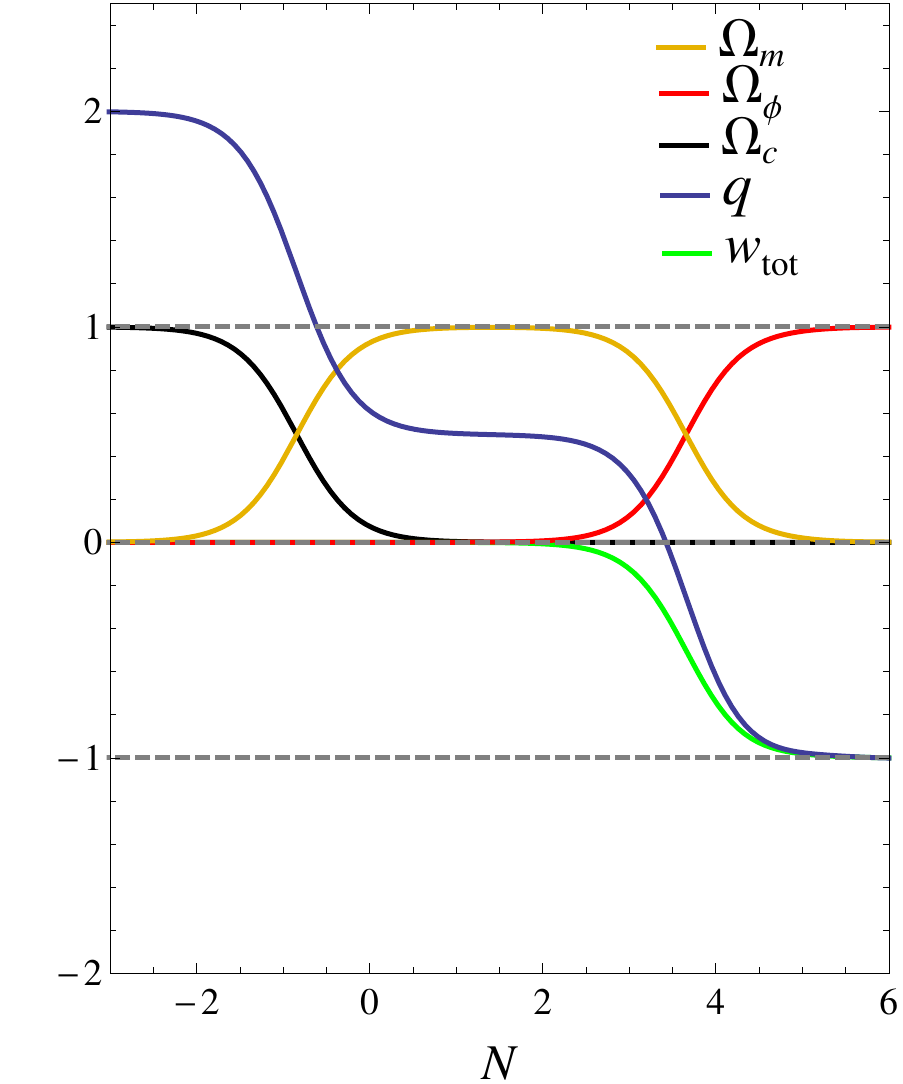}\label{fig:dm_de_phan_II}}
\caption{Plot of $q$, $\Omega_{\phi}$, $\Omega_{\rm m}$ versus $N$ with the potential $V(\phi)=V_0[\cosh(\lambda\phi)]^{-\mu}$. Here we take $\lambda=1$, $\mu=-1$, $\beta=1$ and $\epsilon=1$ in panel (a) whereas $\lambda=1$, $\mu=1$, $\beta=2$ and $\epsilon=-1$  in panel(b).
}
\label{fig:dm_de_sinh_II}
\end{figure}

\subsubsection*{Example 1: $V=V_0\,\cosh^{-\mu}(\lambda\phi)$}

For this potential we have again two copies of critical points $D_2$ and $D_3$, since there are two solutions $s_*=\pm\mu\lambda$. As before, we shall assign $D_i^+$ for $s_*= \mu \lambda$ and $D_i^-$ for $s_*=-\mu \lambda$ ($i=2,3$). Point $D_1$ does not depend on the choice of the potential, while the properties of the other critical points are as follow:

\begin{itemize}

\item {\bf Points $D_2^+$, $D_2^-$}: As discussed in the general case, these points are saddle in nature as its eigenvalue $\lambda_1$ is negative while $\lambda_2$ is positive for any values of $s_*$.

\item {\bf Points $D_3^+$, $D_3^-$}: Since point $D_3$ is always saddle for $s_*>0$ and $s_*<0$ as shown in Fig. \ref{fig:d3_regplot_quin} for general potentials. This implies that points $D_3^+$, $D_3^-$ are saddle in nature.

\item {\bf Set of points $D_4$}: This set corresponds to a late time attractor only if $\beta>-1$ and $\mu<0$, otherwise it is saddle.

\end{itemize}

\noindent
Fig~\ref{fig:weff_sinh_quin_II} shows the evolution of the relevant cosmological parameters given suitable initial conditions.
From this figure, we see that the universe can evolve towards an accelerated phase with $q=-1$ (set $D_4$), with a long lasting matter phase dominated by the interacting energy.
From Fig.~\ref{fig:dm_de_quin_II} instead we see that the universe undergoes a transition to a DE dominated phase (set $D_4$ with $z=0$) from a long lasting matter phase dominated by DM (point $D_1$).
In both cases, we find a long lasting matter phase ($w_{\rm tot} = 0$) as required for the formation of the cosmic structure, before transition to DE domination prevails.

\subsubsection*{Example 2: $ V=\frac{M^{4+n}}{\phi^n}$ }

We recall that for this potential one finds $s_*=0$.
Again the properties of point $D_1$ are independent of the choice of the scalar field potential, while the other critical points behave as follows:

\begin{figure}
\centering
\subfigure[]{%
\includegraphics[width=6cm,height=4cm]{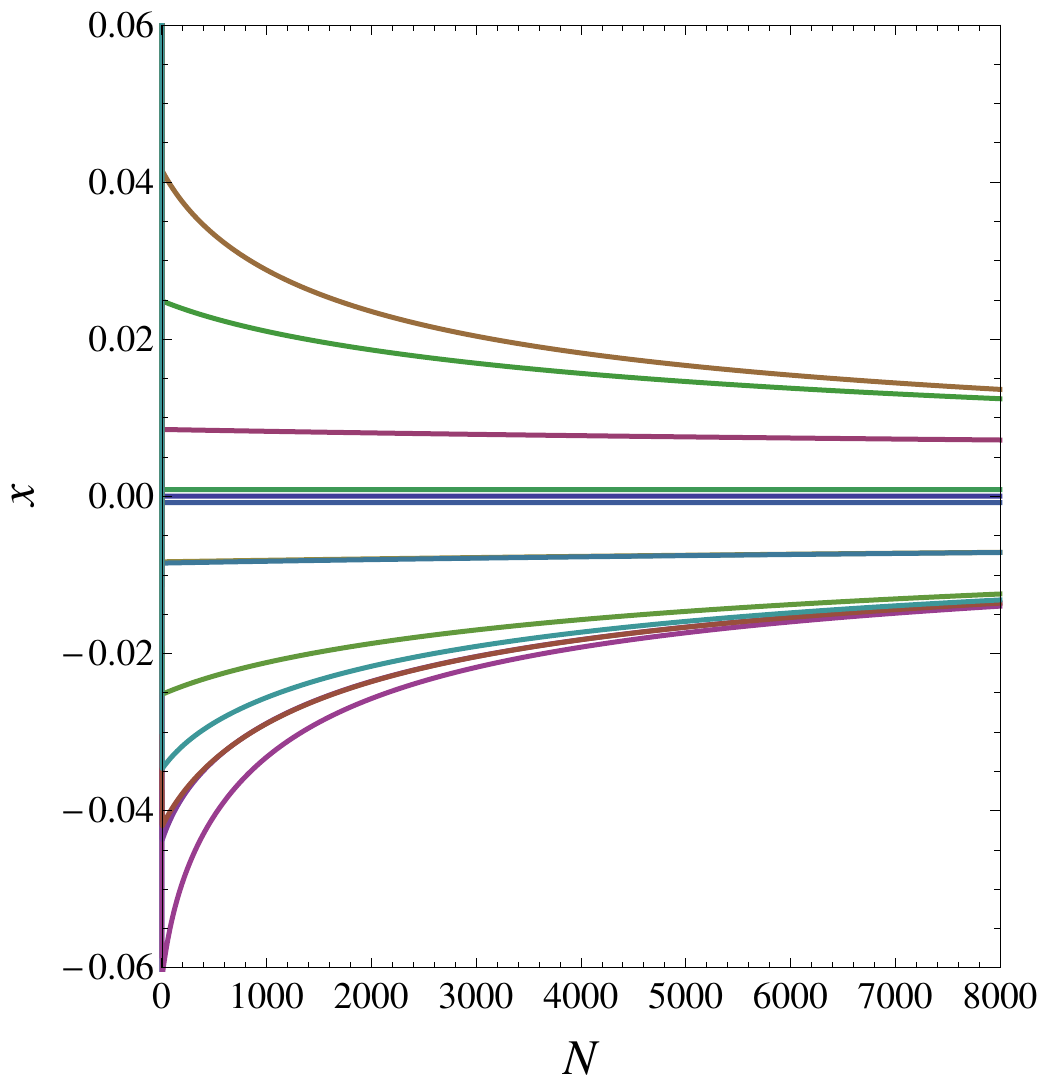}\label{fig:perturb_d3_powerlaw_x}}
\qquad
\subfigure[]{%
\includegraphics[width=6cm,height=4cm]{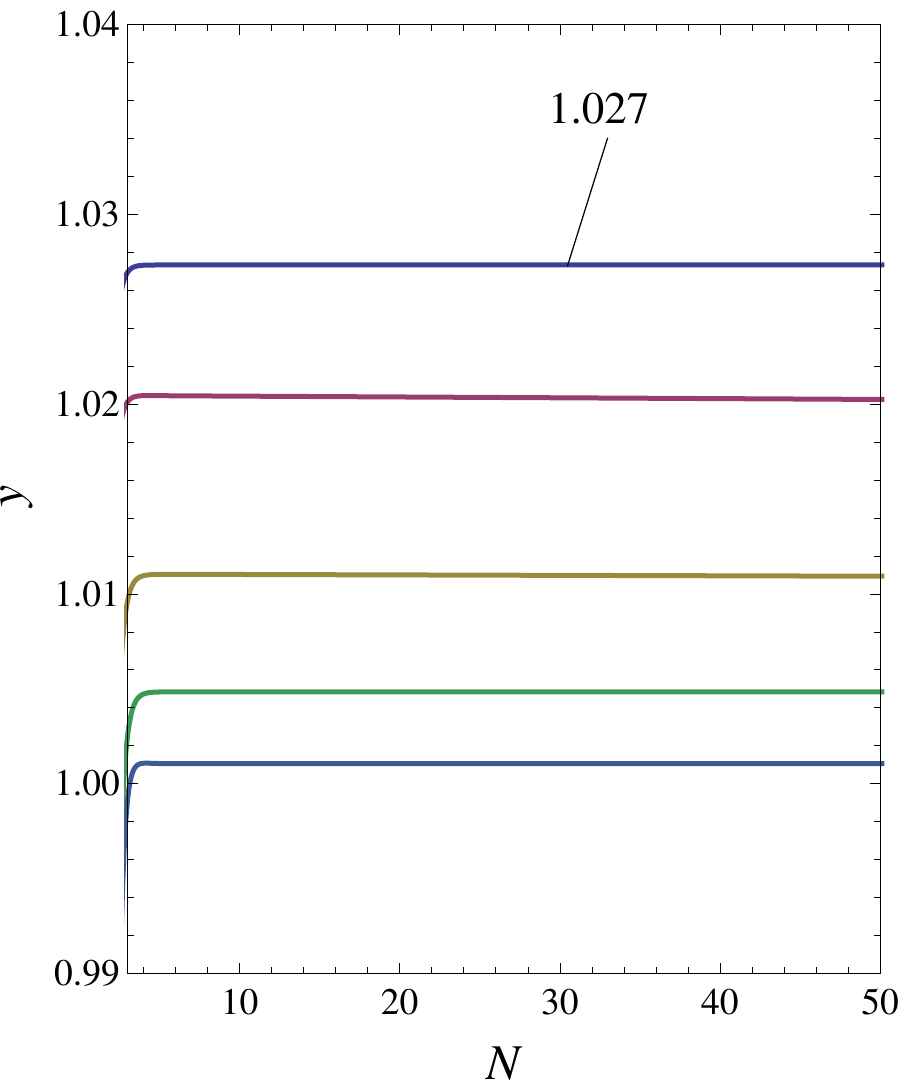}\label{fig:perturb_d3_powerlaw_y}}
\qquad
\subfigure[]{%
\includegraphics[width=6cm,height=4cm]{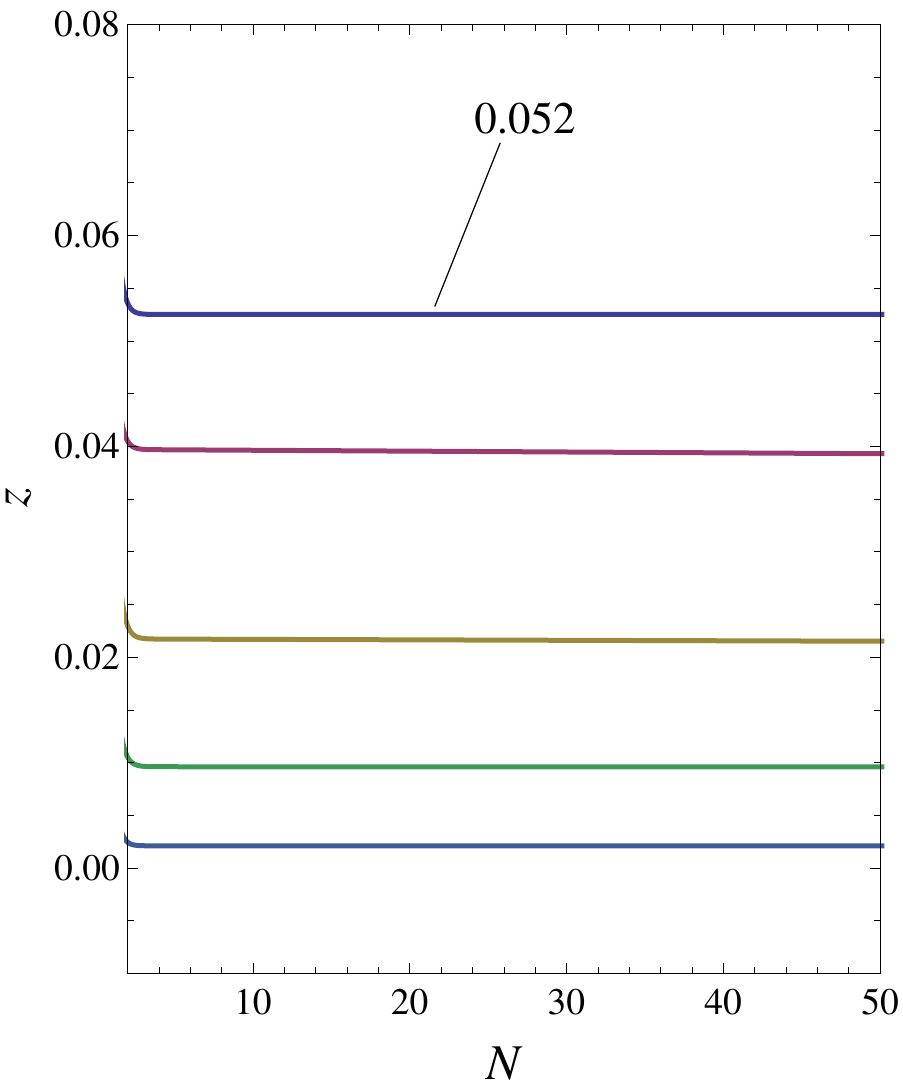}\label{fig:perturb_d3_powerlaw_z}}
\qquad
\subfigure[]{%
\includegraphics[width=6cm,height=4cm]{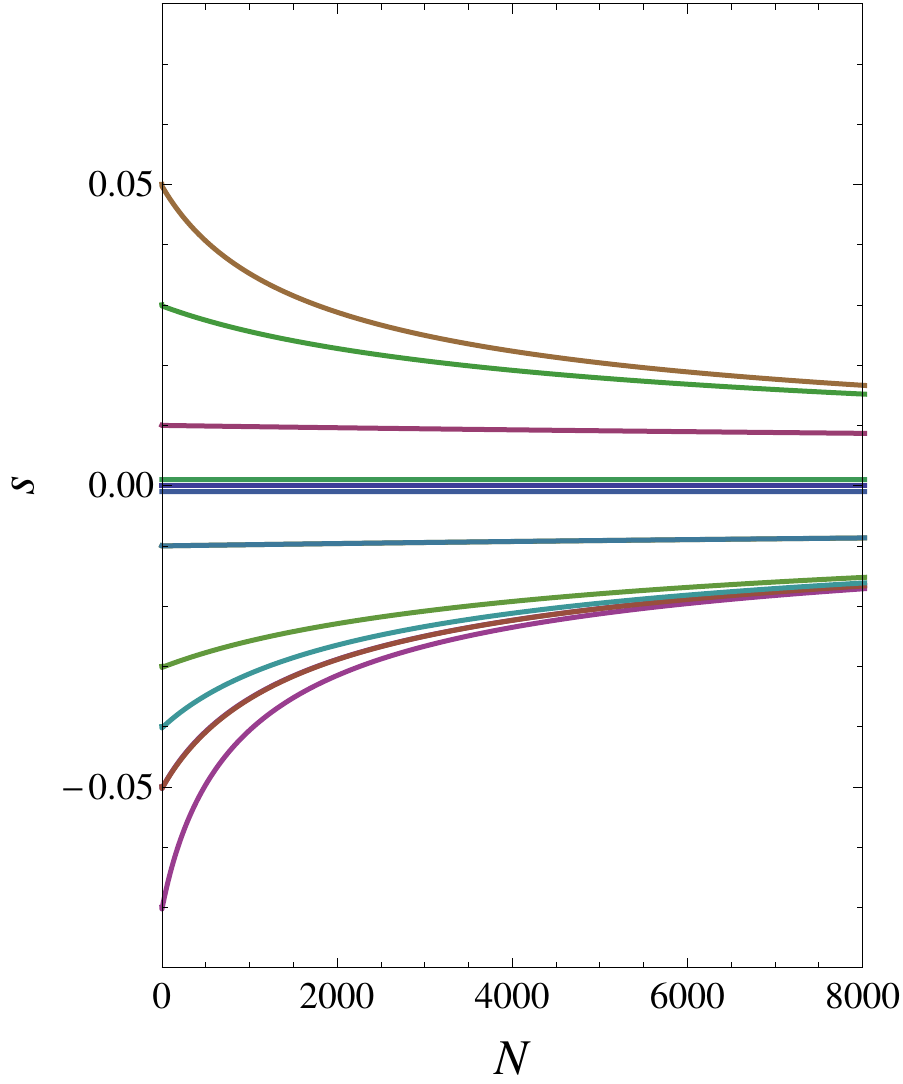}\label{fig:perturb_d3_powerlaw_s}}

\caption{Evolution of trajectories approaching point $D_4$ projected on the $x$, $y$, $z$ and $s$ axis, respectively. Here we considered the potential $ V(\phi)=\frac{M^{4+n}}{\phi^n}$ with $\beta=1$, $n=10$, $\epsilon=1$.
}
\label{fig:perturb_d3_powerlaw}
\end{figure}

\begin{itemize}

\item {\bf Point $D_2$}: This point is non-hyperbolic but it behaves as a saddle since its eigenvalues $\lambda_1$ and $\lambda_2$ are of opposite sign.

\item {\bf Point $D_3$}: This point does not exist for this potential since $s_*=0$.

\item {\bf Set of points $D_4$}: This set of points is no longer
normally hyperbolic since now $f(0)=0$ and two eigenvalues are
zero. The stability can be determined numerically by plotting the
projection of trajectories near the set separately on the $x$, $y$,
$z$, $s$ axes (see for example
Fig.~\ref{fig:perturb_d3_powerlaw}). We find numerically that
trajectories approach points lying on this set only for $n>0$ and for any choice of $\beta$.
This set can thus behave as a late time attractor.

\end{itemize}

\subsection{\bf Phantom dark energy $(\epsilon=-1)$}

We turn now our attention to phantom DE.
The properties of critical points in table~\ref{Table4} with $\epsilon = -1$ are as follow:

\begin{itemize}

 \item {\bf Point $D_1$}: As in the case of quintessence, this point is always saddle.

\item {\bf Point $D_2$}: This point corresponds to an accelerated universe (if $\beta>2$) with negative DE density. It is stable when $s_*\sqrt{\frac{\beta-1}{\beta+1}}>\frac{6}{\beta+1}$, otherwise it is saddle.

\item {\bf Point $D_3$}: As in the case of quintessence, the complicated stability of this scaling solution can only be determined numerically by plotting the region of negativity of eigenvalues $\lambda_1$ and $\lambda_3$ on the $(s_*,\beta)$ parameter space as shown in Fig.~\ref{fig:d3_regplot_phan}.
This point is again always saddle since the region where $\lambda_1<0$ and $\lambda_3<0$ are disjoint.

\item {\bf Set of points $D_4$}: This set corresponds to a late time attractor only if $\beta>-1$ and $f(0)<0$, otherwise it is saddle. Further investigation is required when $f(0)=0$ and we shall postpone this till a particular potential is chosen.

\end{itemize}

\noindent
The dynamics of this particular model is simple if compared to the interacting model I investigated in section~\ref{section3}.
From the above analysis, we see that the universe can evolve from a matter dominated phase (point $D_1$) towards an accelerated scalar field dominated phase (set $D_4$).
This might explain the late time transition of the universe from a matter dominated phase to a DE dominated phase.
However since the existence and stability behavior of points $D_2$, $D_3$ and $D_4$ depend on $s_*$, $df(s_*)$ and $f(0)$, we shall better analyze their properties for the two potentials considered in our examples.

 \subsubsection*{Example 1: $V=V_0\,\cosh^{-\mu}(\lambda\phi)$}

We recall again that for this potential we obtain $s_* = \pm \mu \lambda$, implying that two copies of points $D_2$ and $D_3$ are present in the phase space.
As before point $D_1$ does not depend on the particular form of the scalar field potential and its properties do not change.
For the other critical points we have:

\begin{itemize}

\item{\bf Points $D_2^+$, $D_2^-$}: Point $D_2^+$ is stable when $\sqrt{\frac{\beta-1}{\beta+1}}>\frac{6}{\beta+1}$ and point $D_2^-$ is stable when $\sqrt{\frac{\beta-1}{\beta+1}}<\frac{6}{\beta+1}$, otherwise they are saddle.

\item {\bf Points $D_3^+$, $D_3^-$}:  As we have seen in the general case, these scaling solution are always saddle for $s_*>0$ and $s_*<0$ (see Fig. \ref{fig:d3_regplot_phan}). This implies that points $D_3^+$, $D_3^-$ are saddle in nature.

\item {\bf Set of points $D_4$}: This set corresponds to a late time attractor  only if $\beta>-1$ and $\mu>0$, otherwise it is saddle.

\end{itemize}

\noindent
The phantom model can yield an interesting and particular dynamics including finite stages of phantom domination.
An example is given in Fig.~\ref{fig:weff_sinh_phan_II} where the universe undergoes a transition from a matter phase dominated by the interacting energy, to a DE final era characterized by oscillations of $w_{\rm tot}$ which bounces the universe between quintessence and phantom domination before eventually stabilising to an effective cosmological constant behavior ($w_{\rm tot} = -1$).
The same model can however leads to a dynamics very similar to standard $\Lambda$CDM, as reported in the example in Fig.~\ref{fig:dm_de_phan_II}.
A long lasting matter dominated phase is in fact followed by a smooth transition to DE domination mimicking the effects of a cosmological constant.

 \subsubsection*{Example 2: $ V=\frac{M^{4+n}}{\phi^n}$ }

\begin{figure}
\centering
\subfigure[]{%
\includegraphics[width=6cm,height=4cm]{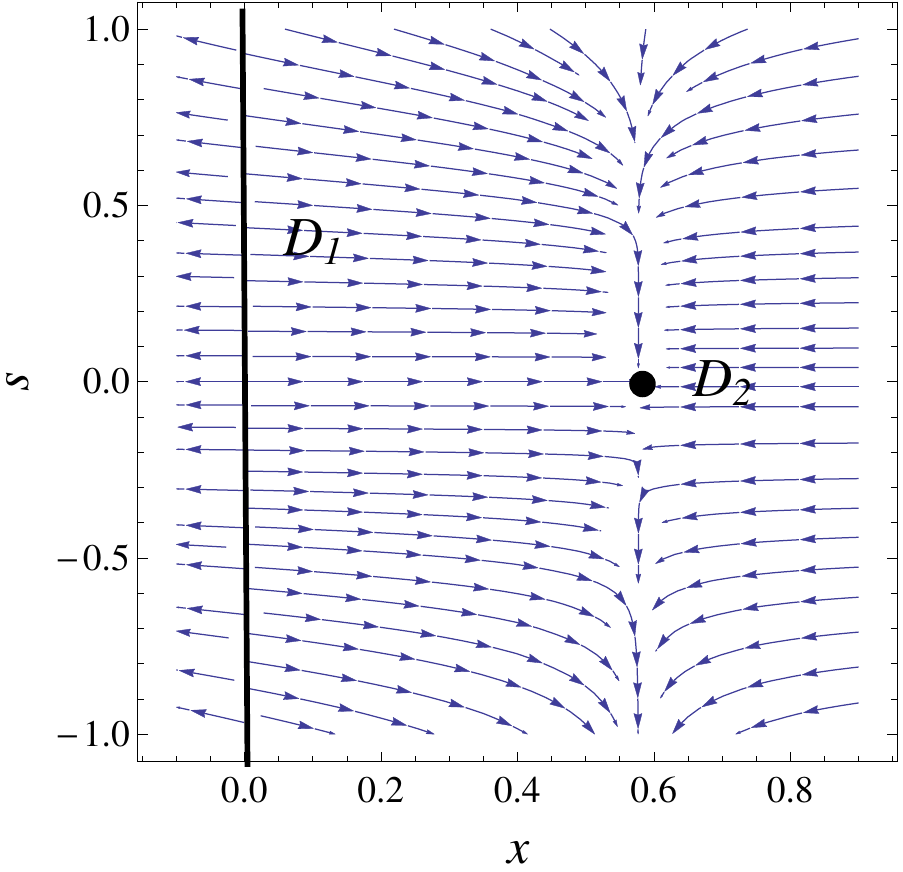}\label{fig:strm_d3_powerlaw_phan_II}}
\qquad
\subfigure[]{%
\includegraphics[width=6cm,height=4cm]{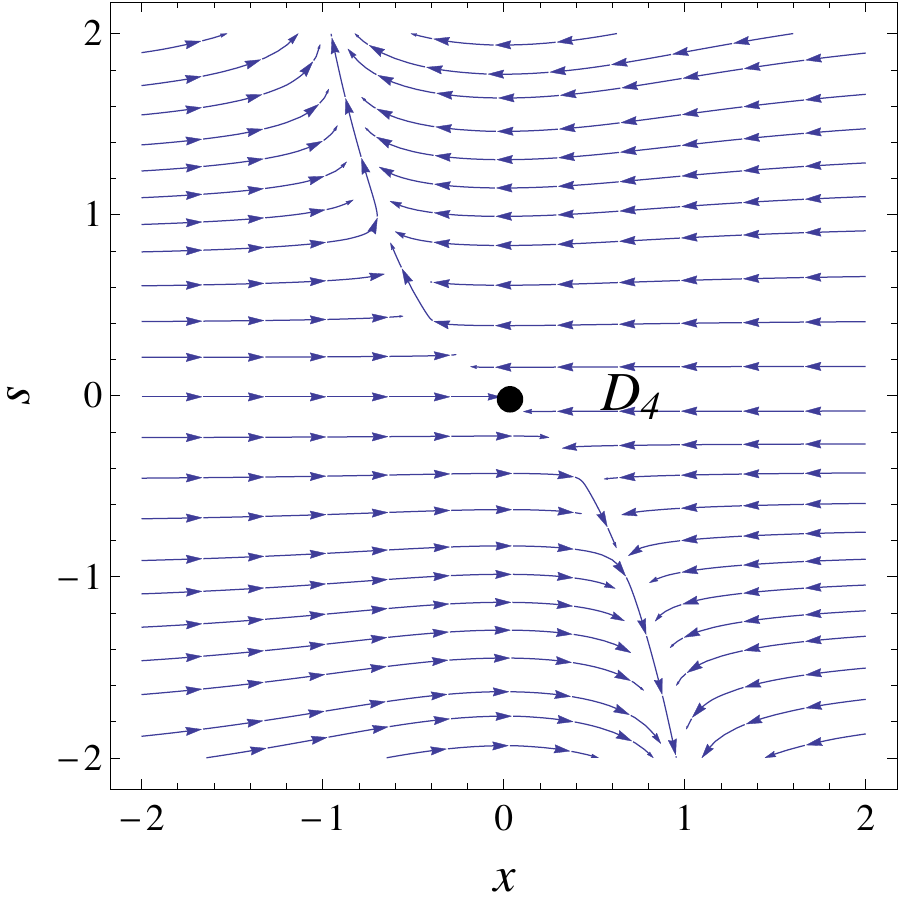}\label{fig:strm_d4_powerlaw_phan_II}}
\caption{(a). Stream plot projection of the dynamical flow of the system (\ref{x_II})-(\ref{s_II}) on the $(x,s)$ subspace. (b) Stream plot projection of the dynamical flow of the system (\ref{x_II})-(\ref{s_II}) on the $(x,1,0,s)$ phase plane. Here we take the potential $ V(\phi)=\frac{M^{4+n}}{\phi^n}$ with $\beta=2$, $n=10$, $\epsilon=-1$. 
}
\label{fig:strm_d3_d4_powerlaw_phan_II}
\end{figure}

For this example we have again $s_* = 0$.
The properties of the critical points are then:

 \begin{itemize}

\item{\bf Point $D_2$}:
This is now a non-hyperbolic point with two vanishing eigenvalues.
Its stability can be determined numerically by plotting the stream plot trajectories on the $(x,s)$ plane.
In fact no matter the values of the model parameters, point $D_2$ is always a saddle in the $(x,s)$ plane, implying that it can never be stable in general, as shown for example in Fig.~\ref{fig:strm_d3_powerlaw_phan_II}.

\item {\bf Point $D_3$}:  This point does not exists since for this potential $s_*=0$.

\item {\bf Set of points $D_4$}:
Again the stability of this set of points is verified numerically.
Independently from the values of the model parameters this set of critical points never represents a stable attractor, as it can be seen in the example of Fig.~\ref{fig:strm_d4_powerlaw_phan_II}.

\end{itemize}

\noindent
We can conclude that the phantom field with power-law potential has no finite late time attractor, implying that the final state of the universe is characterized by some critical point at infinity.
This is in contrast with the example above for the hyperbolic potential, where both points $D_2$ and $D_4$ could be stable.

\section{Cosmological Implications} 
\label{sec:cosmological_implications}

In this section we extract the cosmological features obtained from the interacting DE models analysed above.
The cosmological dynamics of both models is full of phenomenologically interesting solutions and it moreover shows the contribution of loop quantum gravity corrections in both the late and early time behaviors of our universe. 
In what follows we discuss each physically relevant solution separately:

\begin{itemize}

	\item \textbf{Late-time DE dominated solutions}: These critical points are late time attractors characterized by a cosmic phase dominated by DE ($\Omega_\phi = 1$) with an effective EoS mimicking a cosmological constant behavior ($w_{\rm tot} \simeq -1$). Although this kind of solutions naturally arise for quintessence in EC, they do not appear explicitly in the two models investigated here. There are however other solutions which well describe this cosmological behavior. For example in model~I the set of points $C_6$ well represent a late time attractor with $w_{\rm tot} = -1$, as also shown in the example with an hyperbolic potential (see Fig.~\ref{fig:weff_sinh_quin_I}). The same situation emerges in model~II where the set of point $D_4$ acts as a late time attractor mimicking a cosmological constant behavior for both the quintessence and phantom fields (see Figs.~\ref{fig:weff_sinh_quin_II} and \ref{fig:dm_de_sinh_II}). In all these cases a smooth transition from a matter dominated phase to an effective DE era is attained in agreement with the observed dynamics of our universe.

	\item \textbf{Accelerating scaling solutions:} These solutions are identified by a constant finite ratio between $\Omega_\phi$ and $\Omega_m$, implying that the DM energy density evolves at the same rate of the DE energy density. A scenario where a late time attractor is characterised by an accelerating scaling solution is commonly used to alleviate the cosmic coincidence problem, especially in models of interacting DE \cite{Amendola:1999er}. In the models analysed here several scaling solutions appear. For example critical points $C_2$ and $C_3$ in model~I, or point $D_2$ in model~II. Depending on the choice of the parameters, all these points can describe a stable accelerating scaling solution for either quintessence or phantom DE. Point $C_3$ is of particular interest since not only $\Omega_\phi$ scales as $\Omega_m$, but also the contribution due to loop correction $\Omega_c$ scales accordingly remaining a non negligible fraction of the energy content in the universe even at late times.

	\item \textbf{Phantom behavior}: The phantom regime is associated with $w_{\rm tot}< -1$, which can indeed be attained by phantom DE in EC. In such cases however the final state of the universe is a big rip singularity at some finite time in the future. Loop quantum corrections are known for being able to avoid this singularity. An example of this situation is provided in Figs.~\ref{fig:weff_sinh_phan_I} and \ref{fig:weff_sinh_phan_II} where phantom domination is turned into cosmological constant-like behavior at late times. Within this scenario the big rip is avoided and the universe eventually reach an expanding de Sitter solution. Nevertheless this leads to a finite period of phantom regime which characterizes the dynamics of the universe after a standard period of matter domination.

	\item \textbf{Loop quantum effects}: Loop quantum corrections are important whenever the term $z = \rho / \rho_c$ is not negligible. This always happens at early times when $\rho \gg \rho_c$ and $\Omega_c$ dominates over the energy budget of the universe (see Figs.~\ref{fig:weff_sinh_I}, \ref{fig:weff_sinh_II} and \ref{fig:dm_de_sinh_II}). Note that although $\Omega_c$ dominates the effective EoS $w_{\rm tot}$ is either zero or one, implying that the universe at early times is always well described by either a matter dominated phase or a stiff fluid dominated phase. On the other hand late time loop quantum effects do also appear. These are described by the sets of critical points $C_6$ and $D_4$, which can be used to describe the late time accelerated expansion of the universe, mimicking a cosmological constant behavior. For these solutions the DE energy density $\Omega_\phi$ scales according to the contribution of quantum correction $\Omega_c$, similar to a sort of scaling solution (see e.g.~Fig.~\ref{fig:weff_sinh_quin_II}).

	\item \textbf{Possible observational signatures}: The models investigated in this paper present also some other particular features which might provide distinguishing observational signatures with respect to the standard $\Lambda$CDM model. Clear examples are shown in Figs.~\ref{fig:weff_sinh_phan_I} and \ref{fig:weff_sinh_phan_II}, where the effective EoS $w_{\rm tot}$ instead of smoothly changing from zero to $-1$ presents particular oscillations with excursions in the phantom regime. This scenario could in fact be constrained by future data once astronomical observations will better determine the dynamics and EoS of DE at higher redshift. From Fig.~\ref{fig:weff_sinh_quin_I} we note also that early DE might appear in the dynamics of model~I. These scenarios might provide distinguishing observational features, especially at the perturbations level, with respect to $\Lambda$CDM, and can thus in principle be constrained by present and future observations.

\end{itemize}

\smallskip

\section{Conclusion}
\label{section5}

In this paper, we studied the dynamics of interacting quintessence and phantom DE in the LQC framework.
The scope was to perform a complete dynamical system investigation of interacting DE with an arbitrary scalar field potentials.
We have considered two specific interactions between DE and DM of the form $\alpha\,\rho_m\,\dot{\phi}$ (model~I; see Section~\ref{section3}) and $\beta\dot{\rho_{\phi}}$ (model~II; see section~\ref{section4}).
For each of these interactions, we have analysed two forms of dark energy: the quintessence and the phantom fields.
To better characterize the dynamical properties of these interacting models beyond the standard exponential potential, we followed the well known approach of considering the quantity $\Gamma$ (cf.~Eq.~\eqref{eq:gamma}) as a general function of the parameter $s$ (cf.~Eq.~\eqref{7}).
Within this analysis the dimension of the resulting autonomous systems increases from three to four if compared with the corresponding dynamical systems obtained assuming an exponential potential, and the number of critical points multiply making these new systems more difficult to study in detail.
Furthermore in order to better understand the dynamics of these models, especially concerning non-hyperbolic critical points, in each case we considered two concrete non exponential potentials, namely the hyperbolic potential $V=V_0\,\cosh^{-\mu}(\lambda\phi)$ and the inverse power-law potential $ V(\phi)=\frac{M^{4+n}}{\phi^n}$.

We found some interesting and unique features arising from these interacting dark energy models beyond the exponential potential.
In model I with a quintessence field, we obtained one additional set of critical points $C_6$ (see Table~\ref{Table1}), which does not appear in EC.
In fact this set implies the contribution of loop quantum gravity phenomena to a late time accelerated universe.
We also observed in the case of quintessence that there is the possibility of attaining a transient period of phantom acceleration (see Fig.~\ref{fig:weff_sinh_I}), predicting in this way some specific observational signatures of this model.
Furthermore for some choices of the model parameters, we found multiple late time attractors.
This situations are interesting from the mathematical point of view since the dynamics strongly depends on the choice of initial conditions and can be handled using bifurcation theory.
In the case of phantom field, the set $C_6$ is a late time attracting set only for the hyperbolic potential and has no equivalent in the exponential potential case \cite{Fu:2008gh}.
This seems to imply that the scalar field potential plays an important role in determining the contribution of loop quantum effect for driving the late time accelerating universe.
As in the case of exponential potential, we also observe that future big rip singularities can be avoided with the universe passing from phantom domination to an accelerating de Sitter phase ($q=-1$).
On the other hand for power-law potential interacting model~I cannot explain the observed late time behavior of the universe, as there are no late time accelerating attractors.

In the interacting model II with a quintessence field, apart from a standard matter dominated point $D_1$, we obtained one additional late time accelerated set of critical points (set $D_4$) with contributions from loop quantum effects.
We found that the scaling solution $D_3$ is a saddle, which is in contrast with standard EC, where it is a late time, accelerated
scaling solution \cite{Shahalam:2015sja}.
This model can also explain the late time transition from matter domination to a DE dominated phase (see Fig.~\ref{fig:dm_de_sinh_II}).
In the case of phantom field, the set $D_4$ is a late time attracting set only for the hyperbolic potential but not for the power-law potential.
This again suggests the important role of the scalar field potential to obtain a late time accelerated universe in the framework of LQC.
As it happens for model~I, for the power-law potential the interacting model~II cannot explain the late time behavior of the universe as there are no late time accelerating attractors.

In conclusion we found that the results obtained with an exponential potential can not only be recovered for a wider class of potentials, but also new interesting phenomenology appears (see Sec.~\ref{sec:cosmological_implications}).
This is the case for both the quintessence and phantom DE fields, where several interesting solutions have been derived: for example late time DE domination, accelerating scaling solutions, phantom domination, DM to DE transition.
The rich background dynamics obtained from the models investigated in this work may lead to new interesting signatures to look for in present and future observations.
The next logical step to further analyse these models would be to study the dynamics of cosmological perturbations (linear or non-linear) and to compare the results against observational data in order to constrain the theoretical parameters.
Such investigation is beyond the scope of the present paper and will be left for future works.

\acknowledgments

J.D.~is thankful to IUCAA for warm hospitality and facility of
doing research works.
N.T.~acknowledge support from an Enhanced Eurotalents Fellowship and the Labex P2IO.

\end{document}